\documentclass[status=final]{article} 

\usepackage{arxiv}

\RequirePackage{graphicx} 
\RequirePackage[english]{babel} 
\RequirePackage{xspace} 
\RequirePackage{paralist} 
\RequirePackage[shortlabels]{enumitem} 
\RequirePackage{stmaryrd}
\RequirePackage{booktabs}
\RequirePackage{tabularx} 
\RequirePackage{multicol} 

\newcommand{\cf}{cf.\xspace}
\newcommand{\ie}{that is,\xspace}
\newcommand{\eg}{for example,\xspace}

\newcommand{\ies}{i.e.,\xspace} 
\newcommand{\egs}{e.g.\xspace} 
\newcommand{\Eg}{For example,\xspace}

\newcommand{\egnc}{for example\xspace}

\RequirePackage[numbers,sort&compress]{natbib}
\RequirePackage{url} 
\RequirePackage{doi} 
\RequirePackage{cleveref}
\Crefname{paragraph}{Section}{Sections}

\RequirePackage{tikz} 
\usetikzlibrary{arrows,arrows.meta,positioning,calc,patterns} 

\RequirePackage{acro}
\acsetup{
  first-long-format = \itshape,
  format-include-endings = true,
  group-citation = true}

\DeclareAcronym{FM}{short = FM, long = formal method} 
\DeclareAcronym{iFM}{short = iFM, long = integrated formal method} 
\DeclareAcronym{RAS}{short = RAS, long = robots and autonomous systems} 
\DeclareAcronym{RCA}{short = RCA, long = root cause analysis} 
\DeclareAcronym{DSL}{short = DSL, long = domain-specific language} 
\DeclareAcronym{SWOT}{short = SWOT, long = {strengths, weaknesses, opportunities, and threats}} 
\DeclareAcronym{MDE}{short = MDE, long = model-driven engineering} 
\DeclareAcronym{UML}{short = UML, long = {Unified Modelling Language}} 
\DeclareAcronym{SysML}{short = SysML, long = {Systems Modelling Language}} 
\DeclareAcronym{SIL}{short = SIL, long = safety integrity level} 
\DeclareAcronym{ASIL}{short = ASIL, long = automotive safety integrity level} 
\DeclareAcronym{DAL}{short = DAL, long = design assurance level} 
\DeclareAcronym{SC}{short = SC, long = systematic capability} 
\DeclareAcronym{RE}{short = RE, long = requirements engineering} 
\DeclareAcronym{DSE}{short = DSE, long = dependable systems engineering} 
\DeclareAcronym{DA}{short = DA, long = dependability assurance} 
\DeclareAcronym{IT}{short = IT, long = information technology} 
\DeclareAcronym{FDA}{short = FDA, long = Food and Drug Administration} 
\DeclareAcronym{FI}{short = FI, long = formal inspection} 
\DeclareAcronym{MBD}{short = MBD, long = model-based development} 
\DeclareAcronym{ML}{short = ML, long = machine learning} 
\DeclareAcronym{QA}{short = QA, long = quality assurance} 
\DeclareAcronym{SACM}{short = SACM, long = Structured Assurance Case Meta-model} 
\DeclareAcronym{SMT}{short = SMT, long = Satisfiability Modulo Theory} 
\DeclareAcronym{UTP}{short = UTP, long = Unifying Theories of Programming} 
\DeclareAcronym{HCI}{short = HCI, long = human-computer interaction} 
\DeclareAcronym{AI}{short = AI, long = artificial intelligence} 

\RequirePackage{mdframed}
\newmdtheoremenv[hidealllines=true, skipabove=.5em,
innertopmargin=0, backgroundcolor=gray!20]{strength}{Strength}
\newmdtheoremenv[hidealllines=true, skipabove=.5em,
innertopmargin=0, backgroundcolor=gray!20]{weakness}{Weakness}
\newmdtheoremenv[hidealllines=true, skipabove=.5em,
innertopmargin=0, backgroundcolor=gray!20]{opportunity}{Opportunity}
\newmdtheoremenv[hidealllines=true, skipabove=.5em,
innertopmargin=0, backgroundcolor=gray!20]{threat}{Threat}
\newmdtheoremenv[hidealllines=true, skipabove=.5em,
innertopmargin=0, backgroundcolor=gray!20]{remedy}{Remedy}

\renewcommand*{\appendixname}{} 

\usepackage{xwatermark}
\newwatermark[firstpage,color=gray!60,angle=90,scale=0.32,xpos=3.9in,ypos=0]{%
  \href{https://arxiv.org/abs/1812.10103}{\color{gray}{arXiv:1812.10103 [cs.SE]}}}
\newwatermark[firstpage,color=gray!90,angle=0,scale=0.28,xpos=0in,ypos=-5in]{%
  *correspondence: \texttt{mario.gleirscher@york.ac.uk}}

\usepackage{authblk}

\begin{document}
\title{New Opportunities for Integrated Formal Methods$^1$}

\author{Mario Gleirscher}
\author{Simon Foster}
\author{Jim Woodcock}
\affil{Department of Computer Science, University of York, United Kingdom\\
  Deramore Lane, Heslington, York YO10 5GH}

\footnotetext[1]{Partly funded by the Deutsche
  Forschungsgemeinschaft (DFG, German Research Foundation) under the
  Grant no.~381212925.
  \copyright\ 2019. This is the author's version of the work. It is posted here for
  your personal use. Not for redistribution. The definitive version was published
  in ACM Computing Surveys, \url{https://doi.org/10.1145/3357231}.
} 

\maketitle
\begin{abstract}
  Formal methods have provided approaches for investigating software
  engineering fundamentals and also have high potential to improve
  current practices in dependability assurance.  In this article, we
  summarise known strengths and weaknesses of formal methods.  From
  the perspective of the assurance of \ac{RAS}, we highlight new
  opportunities for integrated formal methods and identify threats to
  the adoption of such methods.  Based on these opportunities and
  threats, we develop an agenda for fundamental and empirical research
  on integrated formal methods and for successful transfer of
  validated research to \ac{RAS} assurance.  Furthermore, we outline
  our expectations on useful outcomes of such an agenda.
\end{abstract}

 \keywords{Formal methods \and strengths \and weaknesses \and opportunities \and
  threats \and SWOT \and challenges \and integration \and unification \and research
  agenda \and robots and autonomous systems}

\begin{multicols}{3}
\label{sec:nomenclature}
\footnotesize
\printacronyms
\normalsize
\end{multicols}

\section{Introduction}
\label{sec:introduction}

A plethora of difficulties in software practice and momentous software
faults have been continuously delivering reasons to believe that a
significantly more rigorous discipline of software engineering is
needed~\cite{Jones2003-earlysearchtractable}.
Researchers such as \citet{Neumann2018} have collected plenty of
anecdotal evidence on software-related risks substantiating this
belief.

In \emph{\acl{DSE}}, researchers have turned this
belief into one of their working hypotheses and contributed
formalisms, techniques, and tools to increase the rigour in
engineering workflows~\cite{Jones2003-earlysearchtractable}.  Examples
of activities where formalisms have been brought to bear include
requirements
engineering~\cite[\egs][]{Gunter2000-referencemodelrequirements},
architecture design and verification, test-driven development, program
synthesis, and testing, to name a few.  \emph{\Acp{FM}} have
been an active research area for decades, serving as powerful tools
for \emph{theoretical researchers}.  As a paradigm to be adopted by
\emph{practitioners} they have also been investigated by \emph{applied
  researchers}.  Applications of \acp{FM} have shown their strengths
but also their weaknesses~\cite{Jones2003-earlysearchtractable,
  MacKenzie2001-MechanizingProofComputing}.

Based on observations contradicting the mentioned difficulties,
\citet{Hoare1996-Howdidsoftware} came to ask ``How did software get so
reliable without proof?''  He, and later
\citet{MacKenzie2001-MechanizingProofComputing},
suggested that continuously improved design of programming languages
and compilers, defensive programming, inspection, and testing must
have effectively prevented many dangerous practices.
\citeauthor{MacKenzie2001-MechanizingProofComputing} coined the
``Hoare paradox'' stating that although proof was seldom used,
software had shown to be surprisingly fit for purpose.  However, faced
with software of increasing complexity~(\egs \ac{RAS}),
\citeauthor{MacKenzie2001-MechanizingProofComputing} pondered how long
\citeauthor{Hoare1996-Howdidsoftware}'s question will remain valid.

Indeed, recently we can observe a plethora of difficulties with
\acl{RAS}~\cite{Neumann2018,Koopman2017-AutonomyCarSafety}.  Such
systems are set to be more broadly deployed in society, thereby
increasing their level of safety criticality~\cite{Guiochet2017} and
requiring a stringent regulatory regime.  A successful method for
regulatory acceptance is provided by structured assurance cases, which
provide comprehensible and indefeasible safety arguments supported by
evidence~\cite{Hawkins2015,Habli2010,Kelly1999}.  However, such
assurance cases---whether or not compliant with standards like
IEC~61508\footnote{For: Functional Safety of
  Electrical/Electronic/Programmable Electronic Safety-related
  Systems.} and DO-178C\footnote{For: Software Considerations in
  Airborne Systems and Equipment Certification.}---can be laborious to
create, complicated to maintain and evolve, and must be rigorously
checked by the evaluation process to ensure that all obligations are
met and confidence in the arguments is
achieved~\cite{Rushby2013-LogicEpistemologySafety,
  Greenwell2006-taxonomyfallaciessystem}. Nevertheless, these are
problems that \acp{FM} are designed to overcome.

In spite of the weaknesses of current \acp{FM}, and encouraged by
their strengths, we believe that their coordinated use within
established processes can reduce critical deficits observable in
dependable systems engineering.  \citet{Luckcuck2018} state that
``there is currently no general framework integrating formal methods
for robotic systems''.  The authors highlight the use of what are
called \emph{\acp{iFM}}\footnote{We reuse the term from the homonymous
  conference
  series~\cite{ArakiGallowayTaguchi1999-IntegratedFormalMethods}.}  in
the construction of assurance cases and the production of evidence as
a key opportunity to meet current \ac{RAS} challenges.  Particularly,
computer-assisted assurance techniques~\cite{Wei2019-SACM}, supported
by evidence provided by \acp{iFM}, can greatly increase confidence in
the sufficiency of assurance cases, and also aid in their maintenance
and evolution through automation.  Moreover, the use of modern
\ac{FM}-based tools to support holistic simulation, prototyping,
and verification activities, at each stage of system, hardware, and
software development, can lead to systems that are demonstrably safe,
secure, and trustworthy.

\subsection{Contribution}
\label{sec:contributions}

We investigate the \emph{potentials for the wider adoption of
  integrated formal methods in dependable systems engineering}, taking
\acl{RAS} as a recent opportunity to foster research in such methods
and to support their successful transfer and application in practice.

We analyse the \emph{strengths, weaknesses, opportunities, and
  threats} of using formal methods in the assurance of dependable
systems such as \ac{RAS}.  For this analysis, we summarise experience
reports on formal method transfer and application.

\begin{figure}
  \centering
  \footnotesize
  \begin{tikzpicture}
    [dim/.style={align=flush left},
    tit/.style={align=left,anchor=south west},
    itm/.style={align=center,minimum height=3em,minimum
      width=3cm,fill=gray!20}, lab/.style={align=center}]

    \node[itm,draw,dotted,fill=white] (methgen) at (0,0) {Formal Methods};
    \node[itm] (methspec) at ($(methgen)+(0,-5em)$)
    {integrated\\Formal Methods};
    \node[itm] (appgen) at
    ($(methspec)+(18em,0)$) {Dependable Systems\\Engineering};
    \node[itm,draw,dotted,fill=white] (appspec) at ($(appgen)+(0,-5em)$) {Assurance\\of \ac{RAS}};

    \node[tit] (approach) at ($(methgen.north west)+(0,0)$) {Approach};
    \node[tit,anchor=south] (research) at
    ($(methgen.north)+(9em,0)$) {Empirical Research};
    \node[tit,anchor=west] (application) at ($(research.east)+(1.3,0)$) {Application};
    \node[tit,anchor=north west,align=left] (research) at
    ($(methspec.south west)+(0,-.5em)$) {Fundamental\\Research};
    
    \draw[arrows={-latex},thick,tips=proper]
    (methgen) edge[dotted,sloped,above] node[lab] {applied and\\
      investigated in} (appgen)
    (methspec) edge[sloped,below] node[lab] {applied and\\
      investigated in} (appspec)
    ;
    \draw[arrows={-latex},thick,tips=proper]
    (methspec) edge[right] node[lab] {enhances} (methgen)
    (appspec) edge[right] node[lab] {specialised core subject of} (appgen)
    (appgen) edge node[lab] {achieve wider\\adoption of} (methspec)
    ;
  \end{tikzpicture}
    \caption{\ac{RAS} assurance as an opportunity for the wider adoption of
      \acp{iFM} in dependable systems practice
    \label{fig:refinedapproach}
  }
\end{figure}
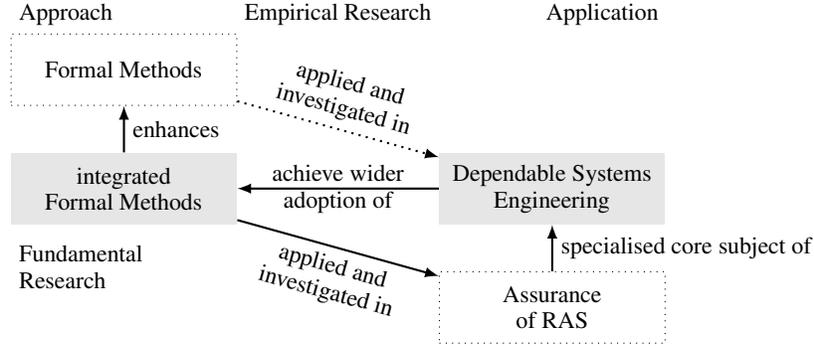

Assuming that
\begin{inparaenum}[(a)]
\item integration enhances \aclp{FM} and 
\item the assurance of \acl{RAS} is a branch of \acl{DSE},
\end{inparaenum}
\Cref{fig:refinedapproach} shows how we derive an
agenda for fundamental and applied \ac{iFM} research.

From the \emph{strengths}, we see in recent research, and from the
\emph{opportunities} in current \ac{RAS} assurance, we argue why this
domain is a key opportunity for \acp{iFM}.  Particularly, we indicate
how such methods can meet typical assurance challenges that \acp{RAS}
are increasingly facing.

From the \emph{weaknesses}, we observe in recent research, and from the
\emph{threats} \acl{FM} research transfer is exposed to, we derive
directions of foundational and empirical research to be taken to
transfer \acp{iFM} into the assurance of robots, autonomous systems,
and other applications, and to use these methods to their maximum
benefit.

Our analysis
\begin{inparaenum}
\item elaborates on the analysis and conclusions of
  \citet{Hoare2009-verifiedsoftwareinitiative},
\item extends their suggestions with regard to formal method
  experimentation and empirical evidence of effectiveness focusing on
  collaboration between formal method researchers and practitioners,
  and
\item develops a research and research transfer road map, placing
  emphasis on \acp{RAS}.
\end{inparaenum}

\subsection{Overview}
\label{sec:overview}

We provide some background including
terminology~(\Cref{sec:terminology}) and related
work~(\Cref{sec:relatedwork}) in the following.  Then, we carry
through an analysis of strengths, weaknesses~(\Cref{sec:fm}),
opportunities~(\Cref{sec:environment}), and
threats~(\Cref{sec:threats}) of \acp{iFM} for \ac{RAS} assurance.
Based on this analysis, we formulate our
hypotheses~(\Cref{sec:position}), pose research questions based on
these hypotheses, derive a research agenda, and specify some outcomes
we expect from this agenda~(\Cref{sec:agenda}).

\section{Background}
\label{sec:background}

This section introduces the core terminology used throughout this
article, and provides a discussion of other surveys of the \ac{FM} state
of the art and a summary of similar positions and agendas.

\subsection{Terminology}
\label{sec:terminology}

For the sake of clarity among readers of different provenance, we restrict
the meaning of some terms we use in the following and introduce
convenient abbreviations.

We view \emph{robots and autonomous systems} as both dependable
systems and highly automated machines capable of achieving a variety
of complex tasks in support of humans.  We can consider such systems
by looking at four layers: the plant or process composed of the
operational environment and the machine; the machine itself; the
machine's controller, and the software embedded into this controller.
Based on these layers, we treat ``embedded system'' and ``embedded
software'' as synonyms.  Machine, controller, and software can all be
distributed.

By \emph{dependable systems engineering}, we refer to error-avoidance
and error-detection activities in control system and embedded software
development~(\egs according to the V-model).
\citet{Avizienis2004-Basicconceptstaxonomy} devised a comprehensive
terminology and an overview of the assessment and handling of a
variety of faults, errors, and failures.  For critical systems, such
activities are expected to be explicit~(\egs traceable, documented),
to employ best practices~(\egs design patterns), and to be driven by
reasonably qualified personnel~(\egs well-trained and experienced
engineers or programmers).

The need for \emph{dependability} often arises from the embedding of
software into a cyber-physical context~(\ies an electronic execution
platform, a physical process to be controlled, and other systems or
human users to interact with).  \Ac{DA}, or
assurance for short, encompasses the usually cross-disciplinary task
of providing evidence for an assurance case~(\egs safety, security,
reliability) for a system in a specific operational
context~\cite{Kelly1999}.

By \emph{formal methods}, we refer to the use of formal~(\ies
mathematically precise and unambiguous) modelling languages to
describe system elements, such as software, hardware, and
the environment, and the subjection of models written in these
languages to analysis~\cite{Jones2003-earlysearchtractable,
  MacKenzie2001-MechanizingProofComputing}, the results of which are
targeted at assurance~\cite{Clarke1996-Formalmethodsstate,DO-333}.
\acp{FM} always require the use of both \emph{formal syntax} and
\emph{formal semantics}~(\ies the mapping of syntax into a
mathematical structure).  Semantics that allows the verification of
refinement or conformance across different \acp{FM} is said to be
\emph{unifying}~\cite{Hoare1998-UnifyingTheoriesProgramming,
  vanGlabbeek2001-HandbookProcessAlgebra}.  \acp{iFM} allow the
coordinated application of several potentially heterogeneous \acp{FM},
supported by interrelated layers of formal
semantics~\cite{Grieskamp2000-IntegratedFormalMethods,
  Boerger2008-AbstractStateMachines}.

\acp{FM} stand in contrast to \emph{informal methods}, which employ
artefacts without a formal syntax or semantics, such as natural
language descriptions and requirements. In the gap between informal
methods and \acp{FM} there is also a variety of \emph{semi-formal
  methods}, including languages like the \ac{UML} and the \ac{SysML},
whose syntax and semantics have frequently been subject of
formalisation in
research~\cite[\egs][]{Giese2004-informalformalspecifications,
  Wieringa1994-IntegratingSemiformal,
  Breu1997-TowardsformalizationUnified,
  Posse2016-executableformalsemantics}.

\emph{\ac{FM}-based tools} assist in the modelling and reasoning based
on a \ac{FM}.  \Ac{MBD} and \ac{MDE} served many opportunities for
\ac{FM}-based tools to be applied in dependable systems
practice~\cite{Woodcock2009,Bicarregui2009}.\footnote{The SCADE Design
  Verifier~(\url{http://www.esterel-technologies.com}) and the seL4
  microkernel~(\url{http://sel4.systems}) represent good although less
  recent examples.}

We speak of \emph{applied or practical \acp{FM}} to signify successful
applications of \acp{FM} in a practical context, \eg to develop
embedded control software deployed in a commercial product marketed by
an industrial company.  We consider the use of \acp{FM} in research
projects still as \emph{\ac{FM} research}.  \emph{Empirical \ac{FM}
  research} investigates practical \acp{FM}, \egnc using surveys, case
studies, or controlled field
experiments~\cite{Goues2018-BridgingGapResearch}.  We speak of
\emph{\ac{FM} transfer} if \ac{FM} research is transferred into
practice with the aim to effectively apply and practice \acp{FM}.  We
consider \ac{FM} transfer, as discussed below, as crucial for
empirical \ac{FM} research and progress of \ac{iFM} research.

\subsection{Related Work}
\label{sec:relatedwork}

Many researchers have suggested that \acp{FM} will, in one way or another,
play a key role in mastering the difficulties discussed below and in
achieving the desired guarantees~(\egs dependability, security,
performance) of future critical
systems~\cite[\egs][]{Jones2003-earlysearchtractable,
  MacKenzie2001-MechanizingProofComputing}.

Expecting an increased use of \acp{FM} to solve practical challenges in the
mid 1990s, \citet{Clarke1996-Formalmethodsstate} suggested \ac{FM}
integration, tool development, and continuous specialist training to
foster successful \ac{FM} transfer to practice.

In 2000, \citet{Lamsweerde2000} observed a growing number of \ac{FM}
success stories in requirements engineering.  Evaluating several \ac{FM}
paradigms, he outlined weaknesses~(\egs isolation of languages, poor
guidance) to be compensated and challenges to be met towards effective
FM use, particularly their integration into multi-paradigm
specification languages.

Aiming at the improvement of software dependability,
\citet{Jackson2007-Softwaredependablesystems} made several key
observations of recent dependability practice~(\egs lack of evidence
for method effectiveness) leading to a general proposal with broad
implications: rigorous dependability cases with explicit claims, the
support of reuse and evolution, and the selective use of \acp{FM}.
Additionally, these authors provide a number of recommendations to
tool vendors and organisations in education and research.

Also in the mid 2000s, \citet{Hinchey2008-Softwareengineeringformal}
spotted a decline of internet software dependability in the context of
an increased level of concurrency in such systems.  Their observation
was backed by an earlier comparative software/hardware dependability
discussion by \citet{Gray2001-DependabilityInternetera}.
\citeauthor{Hinchey2008-Softwareengineeringformal} highlighted
achievements in \ac{FM} automation enabling an increased use of lightweight
\acp{FM} in ``software engineers' usual development environments''.
Furthermore, they stressed the ability to use several \acp{FM} in a combined
manner to verify distributed~(embedded) systems, avoid errors and,
hence, stop the decline of software dependability.

\citet{Hoare2009-verifiedsoftwareinitiative} issued a manifesto for a
``Verified Software Initiative''.  Based on a consensus of strengths,
weaknesses, opportunities, and threats in the software engineering
community, they proposed a long-term international ``research program
towards the construction of error-free software systems''.  This
initiative aims to achieve its agenda through
\begin{inparaenum}[(1)]
\item new theoretical insights into software development,
\item creation of novel automated \ac{FM} tools, and
\item a collection of experiments and benchmarks.
\end{inparaenum}
In particular, the initiative is driven by a number of ``grand
challenges''~\cite{Woodcock2006}---difficult practical verification
problems that can guide future research.  The experiments have broad
scope, and include a smart cash card~\cite{SS-PRG-126}~(the Mondex
card), a secure entry system\footnote{Tokeneer Project Website:
  \url{https://www.adacore.com/tokeneer}.} (Tokeneer), and a cardiac
pacemaker.\footnote{Pacemaker Formal Methods Challenge:
  \url{http://www.cas.mcmaster.ca/sqrl/pacemaker.htm}.}

Outlining an agenda for \ac{FM} transfer,
\citet{Jhala2012-FormalMethodsFuture}
raised the need for improved benchmarks, metrics, and infrastructure
for experimental evaluation, the need for revised teaching and
training curricula~\cite{Perlis1969-IdentifyingDevelopingCurricula},
and the need for research communities interested in engaging with
practitioners and working on ways to scale \acp{FM} up to large systems and
to increase the usability of \acp{FM}.  The authors specified several
applications with great opportunities for \ac{FM} transfer.

Applied researchers and practitioners interviewed by
\citet{Schaffer2016-WhatHappenedFormal} convey an optimistic picture
of \ac{FM} adoption in practice, highlighting the potentials to improve IT
security, particularly in cyber-physical systems.
\citet{Chong2016-ReportNSFWorkshop} share the view that \acp{FM} are the
most promising approach towards acceptably dependable and secure
systems.  The challenges they list for the security domain are similar
to the challenges we perceive in \ac{RAS} assurance: \ac{FM} integration, sound
abstraction techniques, compositional guarantees, and evidence for
sustainable transfer.

With their survey of \acp{FM} for \ac{RAS} verification,
\citet{Luckcuck2018-FormalSpecificationVerification} identified
difficulties of applying \acp{FM} in the robotics domain and summarised
research results and their limitations.  They conclude
\begin{inparaenum}[(i)]
\item that formalisation remains the most critical and most difficult task,
\item that the surveyed approaches do not provide ``sufficient
  evidence for public trust and certification'', and
\item that \acp{iFM} would be highly desirable if the current lack of
  translations between the most relevant of the surveyed
  techniques~(\egs model checking) could be overcome.
\end{inparaenum}
We complement their observations with a further analysis of the lack
of unification of \acp{FM} and of the missing empirical evidence for the
effectiveness of \acp{FM} and \acp{iFM}.  Additionally, we provide a research
road map.

\section{Strengths and Weaknesses of Formal Methods for Assurance}
\label{sec:fm}
\label{sec:strengthsweaknesses}

Following the guidelines for \ac{SWOT} analysis
by~\citet{Piercy1989-MakingSWOTAnalysis}, we provide an overview of
strengths and weaknesses of \acp{FM}, regarding
\begin{itemize}
\item reputation, proof culture, education, training, and use~(\Cref{sec:strweak:reputation}),
\item transfer efforts~(\Cref{sec:strweak:transfer}),
\item evidence of effectiveness~(\Cref{sec:strweak:evidence}),
\item expressivity~(\Cref{sec:strweak:expressivity}), and
\item integration and coordination~(\Cref{sec:strweak:integration}).
\end{itemize}

\subsection{Reputation, Proof Culture, Education, Training, and Use}
\label{sec:strweak:reputation}

In the guest editor's introduction of the ``50
Years of Software Engineering'' IEEE Software special theme
issue~\cite{Erdogmus2018-50YearsSoftware}, the question
\begin{quotation}
  ``Are formal methods essential, or even useful, or are they just an
  intellectual exercise that gets in the way of building real-world
  systems?''
\end{quotation}
invited us to deliberate on
this topic and summarise its highlights.  Applied researchers have
raised the issue of \emph{limited effectiveness and productivity} of
\acp{FM}, particularly in large practical systems with changing
requirements~\cite{Glass2002,Parnas2010}.
\acp{FM} are known to be \emph{difficult to apply in practice},
and \emph{bad communication} between theorists and practitioners
sustains the issue that \acp{FM} are taught but rarely
applied~\cite{Glass2002}.
In contrast, they are considered to have significant potential to cope
with the toughest recent engineering problems: certifiable \ac{RAS}
assurance~\cite{Luckcuck2018}.

Studying the sociology of proof,
\citet{MacKenzie2001-MechanizingProofComputing} identified three
sources of knowledge about a system's dependability: induction~(\ies
from observation), authority~(\egs expert opinion), and
deduction~(\ies inference from models) which is possibly the most powerful.
Since the beginning of software engineering there has been a debate on
the style of deductive reasoning about programs and on the usefulness
of \acp{FM}.
\citet{Millo1979-Socialprocessesproofs} argued that proof is a social
process.  Long and difficult to read computer-produced verification
evidence cannot be subject to such a process and is not genuine proof.
\citet{Dijkstra1978-politicalpamphletmiddle} countered, albeit not as
a supporter of mechanisation, to change from a personal trust-based culture
of proof to the formalisation of proof steps.
\citet{Fetzer1988-Programverificationvery} doubted that verification
based on a model of the program can yield any knowledge about the
dependability of an implementation of that program.
According to \citet{Naur1994-Proofversusformalization}, it is not the
degree of formalisation making a proof convincing but the way the
argument is organised.
\citeauthor{MacKenzie2001-MechanizingProofComputing} tried to
arbitrate this debate between rigorous proof in ordinary mathematics
and formal mechanised proof.  He suggested that proof assistants have
the potential to use formal
methods~\cite{Jones2003-earlysearchtractable} to the maximum benefit.
\citet{Daylight2013-mathematicallogicprogramming} concluded from a
discussion with Tony Hoare that formalist and empiricist perspectives,
while still causing controversies between research and practice,
complement each other in a fruitful way.

Nevertheless, \acp{FM} have shown to be well-suited to
\emph{substantially improve modelling precision, requirements clarity,
  and verification confidence}.  \ac{FM} applications in requirements
engineering such as the ``Software Cost Reduction''
tool set~\cite{Heitmeyer1995-SCRtoolsetspecifying} even carry the
hypothesis of \ac{FM} cost-effectiveness in its name.
By the 1990s, \ac{FM} researchers had already started to examine \ac{FM}
usefulness with the aim to respond to critical observations of
practitioners~\cite{Hall1990,Bowen1995,Knight1997,Barroca1992,%
  Littlewood1998-usecomputerssafety}.  Some of these efforts
culminated in empirical studies
\cite{Sobel2002,Pfleeger1997-Investigatinginfluenceformal} suggesting
\emph{high error detection effectiveness}, though with some
controversy also caused by employed research designs
\cite{Sobel2003-ResponseCommentsFormal,Berry2003-CommentsFormalmethods}.

\citet[Sec.~3.2, Tab.~3.2]{Jones2011-EconomicsSoftwareQuality} observe
that the combination of \emph{formal\footnote{ ``Formal'' refers to
    the formality of the procedure but does not imply the use of
    formal semantics or mathematical techniques.}  inspection}, static
analysis, and formal testing has been the best approach to defect
prevention with \emph{up to 99\% of accumulated defect removal
  efficiency}.  \acp{FM} can be seen as a rigorous and systematic form
of this approach, though less often applied.  In
\Cref{sec:excurs-relat-betw}, we make a brief excursion to the
relationship between \acp{FM} and formal inspection and try to roughly
estimate the population size of \ac{FM} users.

From two larger surveys, one in the early 1990s
\cite{Austin1993-Formalmethodssurvey} and another one in the late
2000s~\cite{Woodcock2009,Bicarregui2009}, we obtain a more
comprehensive picture of the typical advantages of \ac{FM} use and barriers
to \ac{FM} adoption as seen by practitioners and practical \ac{FM} researchers.
In two recent surveys
\cite{Gleirscher2018-safetysurvey,Gleirscher2018-fmsurvey},
we made two, not necessarily surprising but empirically supported,
observations underpinning the main findings of the former studies:
many practitioners \emph{view \acp{FM} as promising} instruments with
high potential, and would use these instruments to their maximum
benefit, whether directly or through \ac{FM}-based tools.  However,
the beneficial use of \acp{FM} is still hindered by severe
obstacles~(\egs \acp{FM} are considered hard to learn, difficult to
integrate in existing processes, too expensive, prone to invalid
abstractions, difficult to maintain).

\begin{strength}
  \acp{FM} can improve \ac{RAS} modelling, the specification of
  \ac{RAS} requirements, and the automation of \ac{RAS} verification,
  fostering the early detection of systematic errors in \ac{RAS}
  designs.
  Many assurance practitioners perceive \ac{FM} usefulness as
  positive.
\end{strength}
\begin{weakness}
  \acp{FM} have shown to be difficult to learn and apply.  Many
  assurance practitioners perceive the ease of use of \acp{FM} as
  negative.  Moreover, research has been ineffectively communicated in
  \ac{FM} teaching and training.
\end{weakness}

\subsection{Transfer Efforts} %
\label{sec:strweak:transfer}

\acp{FM} can be effective in two ways, \emph{ab-initio}~(\ies before
implementation) and \emph{post-facto}~(\ies after implementation).
The ab-initio use of FMs aims at reducing late error costs through,
\eg formal prototyping, step-wise refinement, formal test-driven
development, crafting module assertions prior to programming, or
formal contract-based development.  Once an initial
formalisation~(\egs invariants) is available, it is argued for
families of similar systems that, from the second or third FM
application onward, the benefit of having the formalisation
outperforms the cumulative effort to maintain the formalisation up to
an order of
magnitude~\cite{Miller1998-industrialuseformal,
  Miller1999-FormalVerificationAAMP, Jackson2007-Softwaredependablesystems}.
This argument also addresses agile settings inasmuch as iterations or
increments refer to similar systems.
The post-facto use of \acp{FM} can occur through knowledge extraction
from existing artefacts and using automated tools such as, \eg formal
or model-based post-facto testing tools or post-facto use of code
assertion checkers \cite{Leino2017-AccessibleSoftwareVerification,
  Kaestner2010-AstreeProvingabsence}.  Overall, the second way of
utilising \acp{FM} is known to be more compatible with everyday
software practice.

Achievements collected by~\citet{Aichernig2003},
\citet{Boulanger2012}, and \citet{Gnesi2013} show that many
researchers have been working \emph{towards successful
  \ac{FM} transfer}.  Moreover, researchers experienced in particular
\acp{FM} draw positive conclusions from \ac{FM} applications,
especially in scaling \acp{FM} through adequate tool support for
continuous reasoning in agile software
development~\cite{Miller2010,OHearn2018}.  Other researchers report
about progress in theorem proving of system software of industrial
size~\cite[\egs][]{Klein2018-Formallyverifiedsoftware} and about
FM-based tools for practical
use~\cite[\egs][]{Bicarregui2009,Peleska2016-IndustrialStrengthModel,
  Kaestner2010-AstreeProvingabsence}.
\ac{MBD} and \ac{MDE} have a history of wrapping \acp{FM} into software
tools to make access to formalisms easier and to help automating
tedious tasks via \acp{DSL} and visual
notations.

\emph{Static~(program) analysis} is another branch where
\ac{FM}-based tools have been successfully 
practised~\cite[\egs][]{Kaestner2010-AstreeProvingabsence}.
However, few static analysis tools are based on \acp{FM} and many
of these tools are exposed to reduced effectiveness because of
\emph{high false-positive rates}, particularly if settings are not
perfectly adjusted to the corresponding
project~\cite{Gleirscher2014-IntroductionStaticQuality}.

Furthermore, the concolic testing technique~\cite{Godefroid2005}, a
post-facto \ac{FM}, has seen multiple successes in
industry~\cite{Kim2011,Godefroid2012}. It exercises all possible
execution paths of a program through systematic permutation of a
sequence of branch conditions inferred by an instrumented concrete
execution. It uses these symbolic execution paths and \ac{SMT} solving to
obtain a series of inputs that exercise the full range of program
paths. It does not depend on a predefined model of the program, but
effectively infers one based on the branch conditions. It can
therefore readily be used on existing program developments, and has
notably been used by Samsung for verification of their flash storage
platform software~\cite{Kim2011}. Indeed, it is a belief of the
authors of this latter work that post-facto methods provide greater
opportunities for adoption of \acp{FM} in industry.

Recently, there have been several developments in the use of \acp{FM}
to assure safety requirements related to \ac{HCI}, and a growing body
of
literature~\cite{FormalHCIBook,Harrison2018-FormalHCI,Oliveria2015-FormalPlasticUI},
particularly relating to certification of commercial medical
devices~\cite{Masci2013-InfusionVerifyFDA,Bowen2017-FormalUI}.  There,
\aclp{FM} have been shown to be successful in facilitating
regulatory acceptance, \egnc with the \ac{FDA} agency in the United
States. \citet{Masci2013-InfusionVerifyFDA} formalised the \ac{FDA}
regulatory safety requirements for the user interface of a
patient-controlled infusion pump, and used the PVS proof assistant to
verify it.  \citet{Bowen2017-FormalUI}, similarly, formally modelled
an infusion pump interface, using the Z notation, and then used the
resulting specification as an oracle to generate test cases.
\citet{Harrison2019-FormalDialysis} performed risk analysis for a new
neonatal dialysis machine, formalised safety requirements, and use
these requirements to analyse and verify the source code.  The
consensus of these authors is that \acp{FM} can greatly reduce the
defects in safety critical \ac{HCI} prior to deployment.  The
techniques are increasingly practically applicable through greater
automation.  Moreover, it is likely that these results have potential
applications in other domains, such as \ac{RAS}, provided they can be
transferred and scale to this greater complexity.

\begin{strength}
  There exist many transfer re-entry points from a range of insightful
  \ac{FM} case studies in industrial and academic labs.  \acp{FM} were
  demonstrated to be a useful basis for many static analysis and
  \ac{MDE} tools.
\end{strength}
\begin{weakness}
  The number of practical~(comparative) case studies using (ab-initio)
  \acp{FM} or \acp{iFM}, particularly on \acp{RAS}, is still too low
  to draw useful and firm conclusions on \ac{FM} effectiveness.
\end{weakness}

\subsection{Evidence of Effectiveness}
\label{sec:strweak:evidence}

Whether used as ab-initio or post-facto tools, \emph{strong evidence}
for the efficacy of \acp{FM} in practice is \emph{still
  scarce}~\cite[\egs][]{Pfleeger1997-Investigatinginfluenceformal} and
more anecdotal
\cite[\egs][]{Aichernig2003,Boulanger2012,Gnesi2013,Schaffer2016-WhatHappenedFormal},
rarely drawn from comparative studies
\cite[\egs][]{Sobel2002,Pfleeger1997-Investigatinginfluenceformal},
often primarily conducted in research labs
\cite[\egs][]{Galloway1998,Chudnov2018}, or not recent enough to
reflect latest achievements in verification tool
research~\cite[\egs][]{Catano2002-FormalSpecificationStatic}.  We
observe that a large fraction of empirical evidence for \ac{FM}
effectiveness can be classified as level 6 or 7 according to
\cite[Tab.~2]{Goues2018-BridgingGapResearch}, \ie too weak to draw
effective conclusions.

\citet[p.~39]{Jackson2007-Softwaredependablesystems} as well as
\citet[Sec.~4.4, p.~220]{Jones2011-EconomicsSoftwareQuality}, two
researchers from the software engineering measurement community,
support this observation.
\citeauthor{Jones2011-EconomicsSoftwareQuality} state that ``there is
very little empirical data on several topics that need to be well
understood if proofs of correctness are to become useful tools for
professional software development as opposed to academic
experiments''.  Moreover, the controversies about proof culture
summarised in \Cref{sec:strweak:reputation} contain little data to
resolve practitioners' doubts.  

\citet{Graydon2015} observed this lack of evidence of \ac{FM}
effectiveness in assurance argumentation.  More generally,
\citet{Rae2010} noticed insufficiently evaluated safety research.
About 86\% of works lack guidance to reproduce results, hence forming
a barrier to the advancement of safety practice.  Although their study
is limited to one conference series, it indicates deficiencies in the
evaluation of \ac{DA} research.
Overall, it is important to understand that the mentioned lack of
evidence and successful transfer produces great opportunities for
further empirical and theoretical \ac{FM} research.

\begin{strength}
  For (comparative) studies of \ac{FM} effectiveness, there are
  several research designs and benchmark examples available from the
  scientific literature.  In \Cref{sec:excurs-relat-betw}, we assess
  the effort and feasibility of corresponding qualitative and
  quantitative studies.
\end{strength}
\begin{weakness}
  \acp{FM} have been suffering from fragile effectiveness and
  productivity in dependability engineering in general.
  There is a lack of convincing evidence of \ac{FM} effectiveness,
  particularly, of ab-initio \acp{FM}.  \ac{RAS} engineering and
  assurance are likely to be affected by these weaknesses.
\end{weakness}

\subsection{Expressivity}
\label{sec:strweak:expressivity}

An often quoted weakness of \ac{MBD}, particularly when applied to
\acp{RAS}, is the ``reality gap''~\cite{Brooks1992,Jakobi1995} that
can exist between a naively constructed model and its corresponding
real-world artefact.  According to~\citet{Brooks1992}, over-reliance
on simulation to test behaviour using naive and insufficiently
validated models can lead to effort being applied to solving problems
that do not exist in the real world. Worse, programs for robotic
controllers developed in a model-based setting may fail when executed
on real-world hardware, because ``it is very hard to simulate the
actual dynamics of the real-world''~\cite{Brooks1992}. This problem is
not only true of simulation, but any form of model-based analysis,
including reasoning in \acp{FM}
\cite{Fetzer1988-Programverificationvery}.

The fundamental problem here is that it is impossible to model the
behaviour of any physical entity precisely~\cite{Lee2018}, unless we
replicate the original. Moreover, as models become more detailed,
their utility decreases and they can become just as difficult to
comprehend and analyse as their real-world counterparts, an
observation highlighted by the famous paradox of
\citet{Bonini1962}. Nevertheless, as statistician George Box said
``all models are wrong but some are useful''~\cite{Box1987}: we must
evaluate a model not upon how ``correct'' it is, or how much detail it
contains, but on how \emph{informative} it is.  According
to~\citet{Lee2018}, the antidote is not to abandon the use of models,
but to recognise their inherent limitations and strengths, and apply
them intelligently to reasoning about a specific problem.  This means
selecting appropriate modelling paradigms that enable specification of
behaviour at a sufficiently detailed level of abstraction, and using
the resulting models to guide the engineering process.

\begin{strength}
  \acp{FM} allow and foster the use of specific abstractions to
  specifically inform engineers of \ac{RAS} properties critical for
  their assurance.
\end{strength}
\begin{weakness}
  The effectiveness of formal models is fragile and can be
  significantly reduced because of uncontrollable gaps
  between models and their implementations.
\end{weakness}

\subsection{Integration and Coordination}
\label{sec:strweak:integration}

Modelling notations usually employ a particular paradigm to abstract
the behaviour of the real-world.  For example, the state-based
paradigm, employed by \acp{FM} like Z~\cite{Spivey89-ZRM},
B~\cite{Abrial1996-B}, and refinement
calculus~\cite{Back1989,Mor1996}, considers how the internal
state of a system evolves, whilst the event-driven paradigm, employed
in process calculi like CSP~\cite{Hoare85-CSP},
CCS~\cite{Milner89-CCS}, and
$\pi$-calculus~\cite{Milner1999-PiCalculus}, considers how behaviour
may be influenced by external interactions. Consequently, individual
formal methods are usually limited to considering only certain aspects
or views of a system's
behaviour~\cite{Paige1997FM-IntegratedFormalMethods,Broy1998-IntegratedFormalMethods},
which can limit their effectiveness when used in isolation.  Many
researchers have therefore sought to overcome this weakness by \ac{FM}
integration~\cite{Paige1997FM-IntegratedFormalMethods,
  Galloway1997-IntegratedFormalMethods,
  Broy1998-IntegratedFormalMethods,Clarke1996-Formalmethodsstate}.

The 1990s saw a large number of works on semantic unification and
method
integration~\cite{Paige1997FM-IntegratedFormalMethods,Galloway1997-IntegratedFormalMethods}.
Theoretical foundations were provided by
\citeauthor{Hehner1984-PredicativeProgramming}, in his seminal work on
semantic unification using the ``programs-as-predicates''
approach~\cite{Hehner1984-PredicativeProgramming,Hehner1990} and
comparative semantics~\cite{Hehner1988}.  At the same time, refinement
calculi were developed~\cite{Morris1987,Back1989,Mor1996} that would
underlie the work on linking heterogeneous notations through
abstraction.  Also, \citet{Woodcock1990-StateConcurrency} explored the
integration of state- and event-based modelling using weakest
preconditions, and several other works on this topic
followed~\cite{Evans1994,Galloway1997-ZCCS,Roscoe1997-StateConcurrency}.
\citeauthor{Hoare1994-UnifiedTheories} proposed a unified theory of
programming~\cite{Hoare1994-UnifiedTheories} that links together the
three semantic styles: denotational, operational, and algebraic. These
developments culminated in
\citeauthor{Hoare1998-UnifyingTheoriesProgramming}'s
\ac{UTP}~\cite{Hoare1998-UnifyingTheoriesProgramming}, a general
framework for integration of semantically heterogeneous notations by
application of \citeauthor{HehnerAndHoare1983}'s
approach~\cite{HehnerAndHoare1983} to the formalisation of a catalogue
of computational paradigms, with links between them formalised using
Galois connections.  This framework enables a definitive solution to
the integration of states and events, along with other computational
paradigms, in the \textsc{Circus} language
family~\cite{Oliveira2009-UTPCircus,Butterfield2009-SlottedCircus,Wei2013-CircusTime}.

Another result of these developments was a number of seminal works on
\ac{FM} integration~\cite{Paige1997FM-IntegratedFormalMethods,
  Galloway1997-IntegratedFormalMethods,Broy1998-IntegratedFormalMethods}.
\citeauthor{Paige1997FM-IntegratedFormalMethods}, inspired by work on
systematic method integration~\cite{Kronlof1993-MethodIntegration},
defined a generic ``meta-method'' that aimed at integration of several
formal and semi-formal methods using notational translations with a
common predicative semantic foundation, which builds on
\citeauthor{Hehner1990}'s work~\cite{Hehner1990}. Meanwhile,
\citet{Galloway1997-IntegratedFormalMethods}, building on their
previous work~\cite{Galloway1997-ZCCS}, likewise proposed the creation
of hybrid \acp{FM} with a multi-paradigm approach.  Moreover,
\citet{Broy1998-IntegratedFormalMethods} proposed that \acp{FM}
should be integrated into the V-Model of 
development~\cite{Hoehn2008-DasVModell} with common semantic
foundations to link the various artefacts across development
steps.

These diverse efforts eventually led to the founding of the \ac{iFM}
conference series in
1999~\cite{ArakiGallowayTaguchi1999-IntegratedFormalMethods}, with the
aim of developing theoretical foundations for ``combining behavioural
and state-based formalisms''.  For the second iteration of the \ac{iFM}
conference~\cite{Grieskamp2000-IntegratedFormalMethods}, the scope
broadened to consider all the different aspects of \ac{FM}
integration, including semantic integration, traceability, tool
integration, and refinement. A few years later, a conference series
was also established for \ac{UTP}~\cite{Dunne2006-UTP}, with the aim
of continuing to develop unifying semantics for diverse notations
within the \ac{UTP} framework.

However, there is as yet no agreed and general methodology for
integrating \acp{FM} that could be applied to
\acp{RAS}~\cite{Luckcuck2018}.  Overall, integration is of particular
pertinence to \acp{RAS}, since such systems are multi-layered and
possess a high degree of semantic heterogeneity.  As
\citeauthor{Luckcuck2018} found, they ``can be variously categorised
as embedded, cyber-physical, real-time, hybrid, adaptive and even
autonomous systems, with a typical robotic system being likely to
contain all of these aspects''.  When we consider \acp{RAS}, we must
consider advanced computational paradigms like real-time, hybrid
computation with differential equations, probability, and rigid body
dynamics. This implies the use of several different modelling
languages and paradigms to describe the different aspects, and
therefore a variety of analysis techniques to assure properties of the
overall system.  Assurance of autonomous systems will certainly
therefore require \acp{iFM}~\cite{Luckcuck2018}.
\Cref{fig:refinedapproach} on page~\pageref{fig:refinedapproach}
summarises this relationship.

\begin{strength}
  \acp{iFM} raise the potential of integration and coordination of
  several \acp{FM} to consistently reason about \ac{RAS} properties
  implemented by a combination of various technologies such as
  software, electronic hardware, and mechanical hardware.
\end{strength}
\begin{weakness}
  There is currently no agreed framework for the integration of
  \acp{FM} that would effectively address the needs in the \ac{RAS}
  domain or similar domains.
\end{weakness}

\section{Opportunities for Integrated Formal Methods}
\label{sec:environment}
\label{sec:opportunities}

This section continues with the \emph{environmental part} of our
\ac{SWOT} analysis.  Several key opportunities for the transfer of
\acp{iFM} arise from ongoing assurance challenges, particularly in
\ac{RAS} assurance and from looking at what other disciplines do to
cope with similar challenges.  In the following, we describe
opportunities stemming from
\begin{itemize}
\item the desire for early removal of severe
  errors~(\Cref{sec:desire:earlyremoval}),
\item the desire to learn from accidents and their root
  causes~(\Cref{sec:desire:understerror}),
\item the desire of assurance to form a mature
  discipline~(\Cref{sec:desire:maturity}), and
\item the desire for adequate and dependable
  norms~(\Cref{sec:desire:problemstandards}).
\end{itemize}

\subsection{The Desire for Early Removal of Severe Errors}
\label{sec:desire:earlyremoval}

Summarising major challenges in automotive systems engineering in
2006, \citet[p.~39]{Broy2006-Challengesautomotivesoftware} indicated
that modelling languages used in practice were often not formalised
and the \emph{desired benefits could not be achieved from semi-formal
  languages}.  Moreover, \emph{software engineering was not well
  integrated} with core control and mechanical engineering processes.
Domain engineers would produce software/hardware sub-systems and
mechanical sub-assemblies in undesirable isolation.
\citeauthor{Broy2006-Challengesautomotivesoftware} referred to a lack
of \acp{iFM} for \emph{overall architecture verification}.

Has the situation changed since then?
In model-centric development in embedded software practice~(\egs based
on \acs{UML} or \acs{SysML}), drawbacks can be significant if methods
and tools are not well integrated or trained personnel are
missing~\cite{Liebel2016-Modelbasedengineering}.  Likely,
\citeauthor{Broy2006-Challengesautomotivesoftware}'s criticism remains
in contemporary automatic vehicle engineering and assurance practice.
In fact, he has a recent, clearly negative, but not pessimistic answer
to this question~\cite{Broy2018-YesterdayTodayTomorrow}.  Moreover,
this view is shared by the Autonomy Assurance International
Programme's discussion of assurance barriers,\footnote{See
  \url{https://www.york.ac.uk/assuring-autonomy/body-of-knowledge/}.}
\ie current challenges in \ac{RAS} assurance.  These barriers~(\egs
validation, verification, risk acceptance, simulation, human-robot
interaction) could be addressed by formal engineering models and
calculations based on such models to be used as evidence in assurance
cases.

\label{sec:desire-model-based}
Model-based assurance~\cite{Habli2010,Hawkins2015} uses system models
to structure assurance cases and represents another opportunity for
formal methods in~(through-life) assurance. Assurance arguments that
are purely informal can be difficult to evaluate, and may be subject
to argumentation
fallacies~\cite{Greenwell2006-taxonomyfallaciessystem}.  Consequently,
there have been a number of efforts to formalise and mechanise
assurance cases, both at the argumentation
level~\cite{Rushby2014-MechAssure,Denney2018} and the evidence
level~\cite{Cruanes2013}.  More recently, in \ac{MDE}, the
\ac{SACM}\footnote{See
  \url{https://www.omg.org/spec/SACM/About-SACM/}.} is a standardised
meta-model that supports both structured argumentation and integration
of evidence from diverse system models~\cite{Wei2019-SACM}. \ac{SACM}
unifies several existing argumentation notations, and also
provides support for artefact traceability and terminology.  It could,
in the future, serve as a crucial component for \acp{iFM} in assurance.

Leading voices from applied software engineering research periodically
mention the role of \acp{FM} as a key technology to master upcoming
challenges in assuring critical software
systems~\cite{Moore2016-FourThoughtLeaders}.
A round table about the adoption of \acp{FM} in IT
security~\cite{Schaffer2016-WhatHappenedFormal}
positively evaluated their overall suitability, the combination of \acp{FM}
with testing, and the achievements in \ac{FM} automation.  The panellists
noticed limitations of \acp{FM} in short-time-to-market projects and in
detecting unknown vulnerabilities as well as shortcomings in \ac{FM}
training and adoption in practice.

However, even for mission-critical systems, high costs from late
defect removal and long defect repair
cycles~\cite{Jones2011-EconomicsSoftwareQuality}, as well as dangerous
and fatal\footnote{\Eg the fatal accident involving a Tesla advanced
  driving assistance system,
  \url{https://www.theguardian.com/technology/2018/jun/07/tesla-fatal-crash-silicon-valley-autopilot-mode-report}.}
incidents indicate that assurance in some areas is still driven by
practices failing to assist engineers in overcoming their challenges.
Moreover, \citeauthor{Neumann1995-ComputerrelatedRisks}, an observer
of a multitude of computing risks, stated that ``the needs for better
safety, reliability, security, privacy, and system integrity that I
highlighted 24 years ago in my book, Computer-Related Risks, are still
with us in one form or another
today''~\cite{Neumann1995-ComputerrelatedRisks,Hoffmann2018-Promotingcommonsense,Feitelson2019-TonysLaw}.

\Eg artificial intelligence software---particularly \ac{ML}
components---has been developed at a high pace and used in many
non-critical applications.  Recently, \ac{ML} components are increasingly
deployed in critical domains.  For verification and error removal,
such software has to be transparent and explainable.  Preferring
verifiable algorithms to heuristics,
\citet{Parnas2017-realrisksartificial} recalled the corresponding
engineering principle: ``We cannot trust a device unless we know how
it works''.  One way to follow this principle and establish
transparency is to reverse engineer~(\ies to decode) the functionality
of an \ac{ML} component even if this is not possible in
general~\cite{Ben-David2019-Learnabilitycanbe}.
\acp{FM} can help extract knowledge and reverse engineer abstractions
of \ac{ML} systems to explain their behaviour.  We might then ask to
which extent the reverse engineered and verified functionality serves
as a substitute for the original \ac{ML} component.

These anecdotes make it reasonable to question current assurance
practice.  Seen through the eyes of assurance, they suggest
that we might again be facing a dependable software engineering
\emph{crisis} similar to the one from the late
1960s~\cite{Broy2018-YesterdayTodayTomorrow,Randell2018-FiftyYearsSoftware}.

\begin{opportunity}
  We as method researchers could learn from this crisis and improve
  the way that \acp{FM} can be effectively coordinated to support
  early error removal in practical engineering processes.  We as
  practitioners could learn from this crisis and improve the way that
  we correctly engineer and certify highly automated systems such as
  \acl{RAS}.
\end{opportunity}

\subsection{The Desire to Learn From Accidents and Their Root
  Causes}
\label{sec:desire:understerror}

In the title of \Cref{sec:desire:earlyremoval}, the word ``severe''
refers to the negative consequences \emph{potentially} caused by
errors we want to remove using \acp{iFM}.  The more severe the potential
consequences of an error, the more critical is its early removal.  The
usefulness of \acp{iFM} thus positively correlates with their support in
the removal of critical errors.  However, the estimation of severity
often also requires the careful study of past field
incidents~\cite{Holloway2008-HowPastLoss}.

We speak of \emph{field incidents} to refer to significant operational
events in the \emph{field}~(\ies the environment a technical system is
operated) which are undesired because of their safety risks and their
severe harmful consequences.  Field incidents range from minor
incidents to major accidents.  It is important to separate the
observed effect, the field incident, from its causes or, more
precisely, from the \emph{causal chains of events} leading to the
observed effect.
Hence, this analysis depends on the considered system
perimeter~\cite[\egs][]{Avizienis2004-Basicconceptstaxonomy}.
Depending on the possibilities of \emph{observation} and the depth
pursued in a \ac{RCA}, a conclusion on a possible cause can result in
any combination of, \eg overall system failure, human error, adverse
environmental condition, design fault, hardware fault, software fault,
or specification error.\footnote{Specification errors are also called
  development failures~\cite{Avizienis2004-Basicconceptstaxonomy} and
  can be seen as flaws in the process of requirements validation.}

There are many databases about field incidents: some are comprehensive
and include \ac{RCA}, others are less detailed, and some are confidential,
depending on the regulations in the corresponding application domain
or industry sector.  Based on such databases, accident research,
insurance, and consumer institutions occasionally provide brief root
cause statistics along with accident
statistics.%
\footnote{See \cite[Sec.~1.1]{Gleirscher2014-BehavioralSafetyTechnical}
  for a list of accident databases from such institutions.}

Accident statistics allow certain predictions of the safety of systems
and their operation, \egnc whether risk has been and will be acceptably
low.  Such statistics are also used in estimations of the amount of
field testing necessary\footnote{\Eg according to ``As Low As
  Reasonably Practicable'' or ``So Far As Is Reasonably Practicable''.
  See \url{http://www.hse.gov.uk/risk/theory/alarpcheck.htm}:
  ``Something is reasonably practicable unless its costs are grossly
  disproportionate to the benefits.''} %
to sufficiently reduce risk~\cite{Kalra2016-DrivingSafetyHow}.

\Eg it is well-acknowledged that inadequately designed and
implemented user interfaces are a significant contributory factor in
computer-related
accidents~\cite{MacKenzie1994-Computerrelatedaccidental,Kun2016-AutonomyCarUI,Koopman2017-AutonomyCarSafety}.
But how and how fast have we arrived at this conclusion and how can we
prevent future such incidents?
Without proper empirical investigation of accident causes, such
statistics are \emph{of little use in decisions on measures for
  accident prevention}~\cite{Hopkins2004-Quantitativeriskassessment},
particularly on improvements of engineering processes, methods~(\egs
\acp{iFM}), and technologies~(\egs \ac{iFM} tools) used to build these
systems.  For this, we require more in-depth, possibly formal,
\acp{RCA} and statistics that \emph{relate error removal by \acp{iFM}
  and incident root causes}.  To this extent, \ac{RCA} is a great
opportunity for the investigation of \ac{iFM} effectiveness.

\citet{MacKenzie1994-Computerrelatedaccidental} reported about
deficiencies of information gathering~(\egs \ac{RCA}) for databases on
computer-related accidents back in the early 1990s. %
He noted that independent, well-known, but confidential databases
might reduce under-reporting and were thus believed to improve safety
culture. %
Reporting has not much improved as observed by
\citet[p.~39]{Jackson2007-Softwaredependablesystems}.
To understand the current \ac{RCA} situation, we
studied %
a sample of 377 reports from open field incident databases~(in
aviation, automotive, rail, energy, and others) finding the
following~\cite{Gleirscher2018-safetysurvey}:
\begin{enumerate}
\item \acp{RCA} in these reports were of poor quality, either because they
  were not going deep enough, economically or technically infeasible, or
  inaccessible to us.
\item Particularly, root causes~(\egs software faults, specification
  errors) were rarely documented in a way that useful information about
  the technologies used~(\egs software) or consequences in the
  development process could be retrieved from the reports.
\item Reports in some sectors~(\egs aerospace, rail, power plants,
  process industry) contain more in-depth \acp{RCA} than others~(\egs automotive)
  because of different regulations.
\item Some sectors operate official databases~(\egs
  NHTSA\footnote{National Highway Traffic Safety Administration,
    \url{https://www.nhtsa.gov}.} and NTSB\footnote{National
    Transportation Safety Board, \url{https://www.ntsb.gov}.} in the
  US transportation sector) and others do not~(\egs German road
  transportation sector).  
\item Our findings suggest that even in domains with regulated \ac{RCA},
  reports in open databases tend to be less informative than reports
  in closed databases.
\item The reports from the automotive industry exhibited a
  relatively small fraction of technology-related errors~(\egs
  software-related errors).
\end{enumerate}
To validate our study and to better understand the context of our
findings, we interviewed eight\footnote{%
  These experts have experience with tool qualification~(1)
  and
  safety-critical systems in aviation~(3), railway~(1),
  automotive~(4), and energy systems \& turbines~(1).  We have to keep
  their names confidential.} safety
experts~\cite{Gleirscher2018-safetysurvey}. %
One finding was that, because of an unclear separation of technologies
and a lack of explicit architectural knowledge, a desirable
classification of root causes is sometimes infeasible.  Hence,
accident analysts often close their reports with a level of detail too
low to draw helpful conclusions.  Additionally, one expert stated that
\emph{the hidden number of software-related or software-caused field
  incidents in dependable systems practice is likely much larger than
  the known number}.  This matches our intuition but we are missing
clear evidence.

\citeauthor{Ladkin2013-RootCauseAnalysis}, a researcher involved in
the further development of IEC~61508, demands regulations to mandate
the use of systematic \acp{RCA}.\footnote{From personal
  communication.}  In support of his view, we believe that rigorous
in-depth \acp{RCA} based on \acp{iFM} can be helpful to gain clarity
about actual root causes.
Again, beyond this undesirable form of late error removal, RCA data
is essential for the \emph{measurement of the effectiveness} of
error removal techniques, particularly \acp{iFM}.

The ``Toyota unintended acceleration'' incident exemplifies the
difficulty of drawing conclusions without using powerful \ac{RCA}
techniques: a first \ac{RCA} concluded that floor mats and sticky
throttle pedals caused a fatal car mishap.  A second \ac{RCA} carried
out by NASA experts and based on \emph{testing and automated static
  analysis} of the control system~(\ies software and hardware) was not
conclusive.  A third \ac{RCA}\footnote{See expert interview by
  embedded software journalist from EE Times in 2013 on
  \url{https://www.eetimes.com/document.asp?doc_id=1319903&page_number=1}.}
based on \emph{code reviews}---we could not find out which level of
formal inspection was used---detected defects in the control software
and safety architecture, demonstrated to be likely the causes of the
accident~\cite{Koopman2014-CaseStudyToyota}.

\begin{opportunity}
  We could invest in \aclp{iFM} for RCA based on standardised data
  recording~(\egs aircraft black boxes), especially important for
  \acp{RAS} where human operators cannot practically perform incident
  response.
  Based on lessons from formal RCA, we could further invest in the
  employment of integrated formal methods in \ac{RAS} assurance and
  certification to prevent field incidents, major product recalls, and
  overly lengthy root cause investigations.
\end{opportunity}

\subsection{The Desire of Assurance to Form a Mature Discipline}
\label{sec:desire:maturity}

In his Turing Award acceptance speech in 1981, Tony
\citeauthor{Hoare1981-EmperorsOldClothes} reviewed type safety
precautions in programming languages and concluded: ``In any
respectable branch of engineering, failure to observe such elementary
precautions would have long been against the
law''~\cite{Hoare1981-EmperorsOldClothes}.

Inspired by this comparison, it can be helpful to look at other
engineering disciplines such as civil, mechanical, or electrical
engineering to identify transfer opportunities for \acp{iFM}.  There,
engineers use \acp{FM} in many of their critical tasks.  However,
nowadays these methods are often hidden behind powerful software tools
usable by qualified professional engineers.  Although type systems,
run-time bounds checking, and other variants of assertion checking
have been frequently used in dependable systems practice, the overall
level of \ac{FM} adoption is still comparatively low.

\Eg even in less critical mechanical engineering domains,
vocationally trained engineers use computer-aided engineering, design,
and manufacturing software.  Whether for designing machine parts for
serial production~(\ies specification) or for calculations~(\egs
dimensioning, force or material flow simulations) for these parts and
their assembly~(\ies for prototype verification), these engineers use
tools based on canonical mathematical models.

Nowadays, drawings from computer-aided mechanical design carry at
least two types of semantics: one declarative based on calculus for
dimensioning~(1), and one procedural for the synthesis of
Computer-Numerical-Control programs for production machines processing
materials to realise the drawings~(2).  Note that the unifying base of
these two semantics is geometry, a well-studied mathematical
discipline.
Although higher levels of complexity demand more sophisticated
analytical expertise, typically from engineers with several years of
work experience, many tasks can be accomplished by less trained
engineers using the corresponding tools.

Whereas in computer-aided mechanical design both semantics seem to be
used to a similar extent, in \ac{DA} we observe that analogous semantics
are rarely used even if tools are available, and less often we see~(1)
and~(2) being consistently used.
Low adoption might result from the semantics for dimensioning and
production automation being usually less abstract than the semantics
for verification~(1) and synthesis~(2) of computer programs.
Accordingly, Parnas suggests a shift from correctness proof to
property calculation to develop practical formal
methods~\cite[p.~33]{Parnas2010}.

\emph{Patterns} have had a long history in many disciplines.
In mechanical engineering, patterns are better known as \emph{machine
  elements} and are particularly useful in high-reliability
applications.  Machine elements~(and standardised forms thereof) have
a stabilising impact on the outcome of an engineering project.  The
process of element selection and composition can take tremendous
advantage not only from the \emph{reuse} of proven design knowledge
but also from the reuse of complex calculations~(\egs from gear
transmissions, injection moulding tools, skeleton framings).
Moreover, modern tools typically foster the use of \emph{element
  libraries} and \emph{parametric design}.  Importantly, because the
properties of such elements are in many cases well known, calculations
for assemblies~(\ies \emph{compositional} verification) get relatively
easy. However, the higher the required precision of these calculations,
the more expensive is their computation.

These observations are in line with what we know from collaborations
in robotics, like mechatronics, a discipline where many
engineering domains have to play together well:  \acp{FM} are
heavily used for the analysis of robot controllers and for various
kinds of simulations and tests~\cite{Lozano1979,Meng1993}.

Digital circuit engineering is a domain where \acp{FM} such as model
checking have been successfully applied decades ago.  However,
systematic hardware errors, such as Spectre and Meltdown, and the
\emph{unavailability of temporal specifications} of optimised
operations~(\egs branch-prediction and speculative execution)
discontinue the verifiability of recent computer architectures.  This
lack of verifiability of the assumptions~(\egs partitioning,
information flow) about the execution platform complicates the
verifiability of the software~(\egs an operating system) running on
such a platform.\footnote{See blog post on the seL4 microkernel,
  \url{https://research.csiro.au/tsblog/crisis-security-vs-performance/}.}

\begin{opportunity}
  Dependability assurance has not yet successfully adopted \acp{iFM}
  as a vital part of their key methodologies.  If \acp{FM} seem
  relatively well established in other disciplines, we might also be
  able to successfully transfer \acp{iFM} to \ac{RAS} assurance and 
  assurance in other domains.
  Beyond software design patterns, we could benefit from best
  practices in formal specification and development manifested in
  repositories of \ac{FM} patterns.
  Moreover, we could aim at the further unification of established
  \acp{FM} to provide common formal semantics for various domains.
\end{opportunity}

\subsection{The Desire for Adequate and Dependable Norms}
\label{sec:desire:problemstandards}

Dependable systems practice usually includes the transition from what
is called the \emph{system specification} to an artefact called
\emph{system design}.  Typically, \emph{software and hardware
  specifications} are then derived from these two artefacts before
delving into detailed technology choice and development.
\citet[p.~48]{Jackson2007-Softwaredependablesystems} observe that
safety culture in such a framework is more important than strict
standards, but adequate standards and certification regimes can
establish and strengthen the safety culture desirable in dependable
systems practice.

A striking finding in one of our recent discussions
of dependable systems standards~(\egs IEC~61508,
ISO~26262,\footnote{For: Road Vehicles -- Functional Safety.} DO-178C)
is that \emph{normative parts for specification}~(\ies \acl{RE}), for
specification validation~(\ies avoiding and handling requirements
errors), \emph{and for hazard and risk analysis}~(particularly in
early process stages) seem to be \emph{below the state of the
  art}~\cite{Gleirscher2018-safetysurvey,Feitelson2019-TonysLaw},
despite several observations that significant portions~\cite[\egs
44\%,][]{HSE2003} of the \emph{causes} of safety-critical
software-related incidents fall into the category of
\emph{specification
  errors}~\cite{Knight2002-SafetyCriticalSystems,Knight2014-SafetyStandardsNew}.

If one identifies an error in the software specification, do causes of
this error originate from an erroneous system specification?  How are
system, software, and hardware specifications (formally) related?
Control software typically governs the behaviour of a whole machine.
Hence, core components of a software specification might, modulo an
input/output relation~\cite{Parnas1995-FourVariable}, very well govern
the narrative of the overall system specification.  Inspired by the
issue of specification validity as highlighted in
\cite{Jones2003-earlysearchtractable}, do these standards provide
guidance for checking the validity and consistency of core parts of
system, software, and hardware specifications?

The literature provides plenty of evidence of undesired impacts of
specification errors dating back as early as the investigations of
\citet{Lutz1993} and \citet{Endres1975}.  As reported by
\citet{MacKenzie1994-Computerrelatedaccidental}, the 92\% of
computer-related field incidents caused by human-computer interaction
also illustrate the gap between specifications and capabilities of
humans to interact with automation.  Despite these older figures, we
are talking of one of the most critical parts of standards.
Practitioners could expect to receive strong guidance from these
parts.  Moreover, the requirements to show conformance to these parts
should not be vacuous.

Many standards define \emph{specific sets of requirements}~(\ies for
error removal and fault-tolerance) depending on the level of risk a
system~(or any part of it) might cause.  The higher this level,
the more demanding these requirements.  Examples of demanding
requirements include \ac{SIL}~3-4~(IEC~61508),
\ac{ASIL}~C-D~(ISO~26262), \acl{SC}~3-4~(IEC~61508),
\ac{DAL}~A-B~(DO-178C).  Even for the highest such levels the
mentioned standards only ``highly recommend'' but not mandate the use
of \acp{FM}.

\emph{Guidelines} for embedded software development such as
MISRA:1994~\cite{MIRA1994-DevelopmentGuidelinesVehicle}
\emph{recommend} \acp{FM} for \ac{SIL}~4, although MISRA:2004 no
longer includes such information and instead refers back\footnote{This
  might also the case for MISRA:2012 from March 2013.  We are unaware
  of the opposite but were also unable to receive a copy of this
  version.} to MISRA:1994.  As already mentioned, ISO~26262 as the
overriding standard does not go beyond high recommendation of \acp{FM}
for \ac{ASIL}~D.  \citet{Koopman2014-CaseStudyToyota} reported in 2014
that, in the US, car manufacturers are not required to follow 
MISRA and that there are no other software certification
requirements.  This currently also applies to autonomous road vehicles.

In an interesting anecdote, \citeauthor{Ladkin2013-RootCauseAnalysis}
reported on his lack of success in introducing systematic hazard~(and
risk) analysis methodology into normative parts of this
standard~\cite{Ladkin2013-RootCauseAnalysis}.  Moreover, he
mentioned\footnotemark\ unsuccessful attempts to strengthen the role
of \acp{FM} in IEC~61508 and on the ``broken standardisation'' in
assurance practice.  In reaction to that, he proposed the use of
evidently independent peer reviews to ``dampen committee-capture by
big-company bully players''.\footnotetext{See System Safety Mailing
  List message from 4/11/2018,
  \url{http://www.systemsafetylist.org/4183.htm}
  and~\cite{Ladkin2013-StandardsStandardsImproving}.}

Additionally, \citet{Knight2014-SafetyStandardsNew} observed: ``There
is an expectation by the community that standards will embody the best
available technology and that their presentation will allow
determination of conformance to be fairly straightforward.  A
criticism that is seldom heard is that some standards are, in fact,
technically flawed and poorly presented.''  He exemplifies his
critique by several issues with IEC~61508 and RTCA~DO-178B and
suggests to make the meaning of ``conformance [or compliance] with a
standard'' more rigorous.  Particularly, he encourages to replace
\emph{indirect}~(\ies process-related) evidence~(\egs documentation of
specification activities) in assurance cases by \emph{direct}~(\ies
artefact-related) evidence~(\egs unsuccessful checks for presence of
certain specification faults, successful checks for absence of
implementation errors).\footnote{While formal verification serves the
  check of absence of property violations, conventional testing can
  only serve as a check of presence of such violations.}  With the
observation in software quality control that ``there is little
evidence that conformance to process standards guarantees good
products'', \citet{Kitchenham1996-SoftwareQualityElusive} delivered a
reasonable basis for \citeauthor{Knight2014-SafetyStandardsNew}'s
suggestions.

Regarding the integration of dependability approaches and \acp{FM},
\citet{Bowen1993-Safetycriticalsystems} had stated by 1993 that they
``do not know how to combine formal methods assurance with metrics
collected from other techniques such as fault-tolerance''.  Is this
still an issue?  From a practical viewpoint, standards such as, \eg
IEC~61508, ISO~26262, and DO-178C, provide recommendations about
techniques for the reduction of both random hardware failures~(\egs by
fault-tolerance techniques) and systematic hardware and software
failures~(\egs by \acp{FM}, static analysis, and testing).  If
\acp{iFM} can support the combined application of the recommended
techniques and achieve an improvement in practice then we should
really strive to demonstrate this.

We believe that critical fractions of strong direct evidence can be
delivered through the use of \acp{FM}.  In support of
\citeauthor{Feitelson2019-TonysLaw}'s
argument~\cite{Feitelson2019-TonysLaw}, we see a great opportunity
for an assessment of how the corresponding guidelines in these
standards can be extended and aligned with recent results in \ac{FM}
research.

\begin{opportunity}
  We as researchers and practitioners could support the creation of
  adequate state-of-the-art regulations with improved guidance on
  specification construction and validation.  That way, we could
  foster well-certified high-risk software in a time where dangerous
  autonomous machines are about to get widely deployed in our society.
\end{opportunity}

\section{Threats to the Adoption of Integrated Formal Methods}
\label{sec:threats}
\label{sec:obstacles}

This section closes the \emph{environmental part} of our \ac{SWOT}
analysis by identifying threats to \ac{FM} transfer as well as
challenges that arise from alternative or competing approaches taking
the opportunities mentioned in \Cref{sec:opportunities}.  We also
outline remedies to these threats.

The development of effective \acp{iFM} and their successful transfer into
practice can be impeded by 
\begin{itemize}
\item a lack of agreement on a sound semantic base for domain-specific
  and cross-domain \ac{FM}
  integration~(\Cref{sec:misp-unif,sec:tool-integration}),
\item missing support for widely used and established
  tools~(\Cref{sec:tool-integration}),
\item a lack of interest in practical problems on the side of \ac{FM}
  researchers~(\Cref{sec:host-transf-cult}),
\item a lack of incentives for \ac{FM} researchers to engage with current
  practice and for software practitioners to engage with recent
  theoretical results~(\Cref{sec:host-transf-cult}),
\item a bad reputation among practitioners and applied
  researchers~(\Cref{sec:scepticism}),
\item proofs that are faulty or do not
  scale~(\Cref{sec:erroneousproofs}),
\item the quest for soundness overriding the quest for
  usefulness~(\Cref{sec:soundn-usef-autom}).
\end{itemize}
We discuss these threats and barriers in more detail in the following.

\subsection{Difficulties and Misconceptions of Unification}
\label{sec:misp-unif}

According to~\citet{Broy2006-Challengesautomotivesoftware}, the
successes and failures of semi-formal languages~(\egs \acs{UML}, \acs{SysML})
suggest that \acp{FM}, once wrapped in \ac{FM}-based tools, get exposed to the
\emph{quest for a unified syntax}, one main objective of the \ac{UML}
movement in the 1990s.  Rather than a unified syntax, it is more
desirable to have a unified semantics and several well-defined
mappings to domain-specific syntax wherever
convenient~(\Cref{sec:strweak:integration}).  This approach is
occasionally taken up by \acp{DSL} in \ac{MDE}~(\Cref{sec:strweak:transfer}).
\citet{Harel2004} argued that one cannot achieve proper integration of
methods and notations without a unifying semantics.  This argument
carries over to the problem of tool integration as already discussed
in \Cref{sec:strweak:integration} and revisited below.
Particularly, the following challenges apply to \acp{FM} when used in \ac{MDE}:
\begin{enumerate}
\item the maintenance of a single source of information serving in
  the~(automated) derivation of downstream artefacts~(\egs proof
  results, code via
  synthesis)~\cite{Mohagheghi2012-empiricalstudystate}, %
\item a clear mapping between the \ac{DSL} presented to the
  engineer~(using intuitive notation) and the \ac{DSL} semantics serving as
  the basis of formal verification,
\item the embedding of a lean domain-specific formalism into a
  \emph{common} data model~\cite{Broy2010-SeamlessModelBased} suitable
  for access and manipulation by engineers through their
  \emph{various}
  tools~\cite{Gleirscher2007-IncrementalIntegrationHeterogeneous}.
\end{enumerate}
These challenges are complicated by irreducible unidirectionalities
in automated transformations~(\egs model-to-code) limiting the
desirable round-trip engineering~\cite{Stevens2018}~(\ies the change
between views of the same data).

We discussed \ac{SACM}~\cite{Wei2019-SACM} as an assurance \ac{DSL} in
\Cref{sec:desire-model-based}.
Likewise, \emph{architecture description languages}~(\egs the
Architecture Analysis \& Design Language
\cite{Feiler2004-OverviewSAEArchitecture}, EAST-ADL
\cite{Debruyne2004-EASTADLArchitecture}) are \acp{DSL} for overall embedded
system design.
\acp{DSL} can be seen as one shortcut to the still ongoing efforts of
arriving at a reduced version or a variant of \ac{UML} where a
semantics can be defined for the whole
language~\cite[\egs][]{Posse2016-executableformalsemantics}.

At a higher level of abstraction, so-called \emph{architecture
  frameworks}~(\cf ISO~42010, %
\egs the Department of Defense Architecture Framework)
and \emph{artefact and traceability
  models}~\cite[\egs][]{Ramesh2001-TowardsReferenceModels,Whitehead2007-CollaborationSoftwareEngineering}
have been proposed, aiming at the standardisation of specific parts of
the systems and software engineering life-cycle and of the
documentation and data models used there.
These frameworks and models are similar to the models used in product
data/life-cycle management in fields like mechanical or civil
engineering.

To our best knowledge, no cross-disciplinary semantic unification has
been undertaken yet~(see \Cref{sec:strweak:integration}), serving as a
basis for dependable systems engineering.  Although many of these
approaches have not been developed with the aim of formalisation and
the unification of semantics, we believe that this effort has to be
made when developing powerful \acp{iFM}.

\begin{threat}
  The main threat is the lack of agreement on a sound semantic base
  for domain-specific and cross-domain \acp{iFM}.
\end{threat}

\begin{remedy}
  To reduce this threat, \ac{FM} integration and refinement-based
  software engineering could be better aligned with artefact
  models~\cite[\egs][]{Whitehead2007-CollaborationSoftwareEngineering,
    Mendez-Fernandez2010-MetaModelArtefact}.  This alignment may
  foster the unification of formal semantics to strengthen
  traceability among the artefacts and to aid in a variety of change
  impact analysis tasks in the engineering
  process~\cite[\egs][]{Broy2017-logicalapproachsystems,
    Lamsweerde2009-RequirementsEngineeringSystem}.
\end{remedy}

\subsection{Reluctant Integration Culture and Legacy Processes}
\label{sec:tool-integration}

\emph{Tool integration} is about the integration of engineering \acl{IT},
\eg tools for requirements specification, computer-aided software
engineering, and computer-aided mechanical design.  Among the wide variety
of solutions to capture and track model data, the majority deals with
linking or merging data
models~\cite{Gleirscher2007-IncrementalIntegrationHeterogeneous} in
one or another shallow way~(\egs software repositories, data exchange
formats, product/engineering/application data or life-cycle management
systems).

Some tools with sustainable support are heavyweight, making it
difficult to agree on lean model semantics, while others are
proprietary, accompanied with interest in hiding model semantics.  The
surveys of \citet[pp.~102,104]{Liebel2016-Modelbasedengineering},
\citet[p.~104]{Mohagheghi2012-empiricalstudystate}, and
\citet{Akdur2018-surveymodelingmodel} confirm that method and model
integration have not yet been solved in \ac{MBD}, \ac{MDE}, and dependable
systems practice.
Moreover, frequent proposals by
researchers~\cite[\egs][]{Giese2004-informalformalspecifications,
  Breu1997-TowardsformalizationUnified,Posse2016-executableformalsemantics}
to formalise fragments or variants of \ac{UML} and \ac{SysML}
have not yet received wide attention by
practitioners and standardisation authorities.

\ac{DSL}-based integrated development environments~(\egs using Xtext
and Sirius) get close to what is suitable for \ac{FM}-based tools.
Such tools rely on a trusted representation of the formal semantics
integrating the model data.  For successful \ac{iFM} transfer to
assurance practice, tools need to be built on a lean and open central
system model~\cite[\egs][]{Aravantinos2015-AutoFOCUS3Tooling,
  Huber1996-AutoFocustooldistributed}.

An even greater barrier than loosely integrated tools are legacy
\emph{language and modelling paradigms}, an \emph{established tool and
  method market} carried by \emph{legacy stakeholders} and, possibly,
a \emph{neglected continuous improvement of \ac{FM} education and
  training}.

\begin{threat}
  The main threat is discontinuous and disintegrated \ac{FM}
  education, transfer, and tool development.  
\end{threat}

\begin{remedy}
  To reduce this threat, continuous adaptation and improvement of
  education through teaching, of transfer through training,
  application, and feedback, and of tool development through regulated
  interface standards are necessary.
\end{remedy}

\subsection{Reluctant Transfer Culture and Exaggerated Scepticism}
\label{sec:host-transf-cult}

Finally, the vision of introducing \acp{iFM} into assurance practice
might be hindered by a lack of \ac{FM} researchers able or willing to
engage with industrial assurance practice, as diagnosed
by~\citet{Woodcock2009}.  It is certainly hard work to collect
sufficient evidence for \ac{FM} effectiveness in assurance practice
because of intellectual property rights and other legal issues but
also because of a lack of awareness among \ac{FM}
researchers~\cite{Woodcock2009}.  However, for credible method comparison experiments,
\citet{Jones2011-EconomicsSoftwareQuality} recommended a sample of 20
similar projects split into two groups, 10 projects without
treatment~(\ies not using \acp{FM}) and 10 projects with
treatment~(\ies using \acp{FM}) to establish strong evidence~(\ies
evidence of level 5 or above~\cite{Goues2018-BridgingGapResearch}).

\label{sec:scepticism}
Exaggerated scepticism on the side of practitioners and applied
researchers that has piled up over the years might be the
most important barrier to cross.  Early failures to meet high
expectations on \acp{FM} and \ac{FM} transfer might have led to what
can be called an ``\ac{FM} Winter''.  However, we think crossing a few
other barriers first might make it easier to cope with scepticism in
the assurance community and initiate an ``\ac{FM} Spring'', at least in
assurance practice.  The recent successes with \acp{FM}
for certification of commercial medical
devices~\cite{Masci2013-InfusionVerifyFDA,Harrison2019-FormalDialysis},
and the associated burgeoning field of \acp{FM} for
\ac{HCI}~\cite{FormalHCIBook}, are examples that could help to
overcome this scepticism. Moreover, their is potential for transfer of
these results to \ac{RAS} engineering, where the need for safety
assured \ac{HCI} is also
paramount~\cite{Kun2016-AutonomyCarUI,Koopman2017-AutonomyCarSafety}.

\begin{threat}
  The main threat is the reluctance of \ac{FM} researchers to
  regularly engage in transfer efforts combined with an exaggerated
  scepticism of practitioners and with other mechanisms~(\egs lack of
  funding, poorly focused research evaluation, intellectual property
  rights) preventing both sides from engaging in an effective
  bidirectional transfer.  
\end{threat}

\begin{remedy}
  To reduce this threat, building awareness
  among researchers as well as stronger incentives~(\egs regulation)
  to provide continuous transfer funding for \ac{FM} research~(\egs
  not relying on short-term projects like PhD theses) is needed.  A
  good start on the academic side could then be a specific
  standardised repository of \ac{FM} case studies.
\end{remedy}

\subsection{Too Many Errors in Proofs and Failure to Scale}
\label{sec:erroneousproofs}

From the perspective of measurement,
\citet[Sec.~4.1]{Jones2011-EconomicsSoftwareQuality} stated that
``proofs of correctness sound useful, but [i] errors in the proofs
themselves seem to be common failings not covered by the literature.
Further, large applications may have thousands of provable algorithms,
and [ii] the time required to prove them all might take many years''.
For [i], the authors opposed 7\% of erroneous bug repairs to up to
100\% of erroneous proofs, though stating that the latter was based on
an anecdote and there had been little data around.
\citeauthor{Jones2011-EconomicsSoftwareQuality} elaborated an example for
[ii]: Assuming one provable algorithm per 5 function points\footnote{A
  function point is a measure of the conceptual complexity of an IT
  system relevant for the estimation of the amount of work required to
  engineer this system.}  and on average 4 proofs per
day, %
Microsoft Windows 7~(160,000 function points) would have about 32,000
provable algorithms, taking a qualified software engineer about 36
calendar years.  They highlighted that typically only around 5\% of the
personnel are trained to do this work, assuming that algorithms and
requirements are stable during proof time.

\citeauthor{Jones2011-EconomicsSoftwareQuality}'s argument is based on
the view of practical program verification as a largely manual
activity.  However, research in reasoning about programs has devised
promising approaches to proof automation~(\egs bounded model checkers,
interactive proof assistants).  Assessing such approaches,
\citet{MacKenzie2001-MechanizingProofComputing} compared rigorous but
manual formal proof with mechanised or automated proof.
He could not find a case where mechanised proof threw doubt upon an
established mathematical theorem.
\citeauthor{MacKenzie2001-MechanizingProofComputing} observed that
this underpins the robustness argument of
\citet{Millo1979-Socialprocessesproofs} for proofs as manual social
processes rather than lengthy mechanical
derivations~(\cf\Cref{sec:strweak:evidence}).
However, opposing \citeauthor{Millo1979-Socialprocessesproofs}'s view,
\citet[p.~38]{Jones2003-earlysearchtractable} concluded that
structuring of knowledge about computing ideas is the actual issue,
and not the avoidance of mechanisation.  Furthermore,
\citet[p.~39]{Jones2003-earlysearchtractable} mitigated
\citeauthor{Fetzer1988-Programverificationvery}'s issue of the reality
gap in program verification \cite{Fetzer1988-Programverificationvery}
by referring to researchers' continuous efforts in verifying hardware,
programming languages, and compilers in addition to individual
programs.  Finally, in support of
\citeauthor{Jones2011-EconomicsSoftwareQuality}'s concerns,
\citet[p.~38]{Jones2003-earlysearchtractable} highlighted the validity
of specifications, manual and automated proofs rely on, as a
serious issue indeed having received too little
attention from researchers~(\cf\Cref{sec:desire:problemstandards}).

\begin{threat}
  The main threat is the lack of qualified personnel to cope with the
  required amount and type of proof.  
\end{threat}

\begin{remedy}
  We believe, one angle of attacking this threat is the use of proof
  assistants \emph{well-integrated} with common environments for
  requirements specification and software development combined with
  continuous research and education in the corresponding methods and
  tools.
\end{remedy}

\subsection{Failure to Derive Useful Tools}
\label{sec:soundn-usef-autom}

Being loosely related to erroneous proofs, the \emph{information
  overload through false-positive findings of errors} is a well-known
problem in static program analysis.  Semi-formal pattern
checkers,\footnote{Also called bug finding tools.}
such as PMD and FindBugs, are exposed to this
threat~\cite{Gleirscher2014-IntroductionStaticQuality}.  Additionally,
FM-based verification tools, such as Terminator and
ESC/Java~\cite{Flanagan2002-Extendedstaticchecking},
can be unable to correctly report all potential problems, because they
are bounded and therefore unsound.  While such tools can be helpful,
confronting developers with many irrelevant findings~(\ies false
positives) or a high risk of critical misses~(\ies false negatives)
can lead to decreased use of \ac{FM}-based tools.

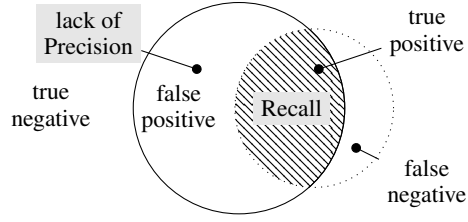
\begin{figure}
  \centering
  \footnotesize
  \begin{tikzpicture}
    [dim/.style={align=flush left},
    tit/.style={align=left,anchor=south west},
    itm/.style={align=center,minimum height=3em,minimum
      width=3cm,fill=gray!20},
    lab/.style={align=center}]

    \draw[] (0,0) circle [radius=40pt];
    \draw[dotted] (1,0) circle [radius=30pt];
    \begin{scope}
      \clip (1,0) circle [radius=30pt];
      \draw[fill=black!20,pattern=north west lines] (0,0) circle [radius=40pt];
    \end{scope}
    
    \node[lab] (TP) at (2.5,1) {true\\positive};
    \node[lab] (TN) at (-2.5,0) {true\\negative};
    \node[lab] (FP) at (-.8,0) {false\\positive};
    \node[lab] (FN) at (2.5,-1) {false\\negative};
    \node[lab,fill=black!10] (prcn) at (-2,1) {lack of\\Precision};
    \node[lab,fill=black!10] (recl) at (0.7,0) {Recall};
    \draw[-{Circle[]}] (prcn) -- (-0.5,0.5);
    \draw[-{Circle[]}] (FN) -- (1.5,-.5);
    \draw[-{Circle[]}] (TP) -- (1,.5);
  \end{tikzpicture}
  \caption{Precision and recall versus soundness and completeness in
    program verification
    \label{fig:precisionrecall}}
\end{figure}

\Cref{fig:precisionrecall} relates the two information retrieval
metrics, \emph{precision} and \emph{recall}, with two adequacy
criteria of proof calculi, \emph{soundness} and \emph{completeness}.
Precision denotes the ratio of actual~(solid circle) to correct~(true
positive findings in the hashed area) findings, while recall measures
the ratio of correct to all theoretically correct~(dotted circle)
findings.  The ideal value for both metrics is 1, meaning that there
are only true negatives and true positives.  Completeness, although
unachievable for richer theories, would correspond to recall and
soundness would correspond to a precision of~1.

On the one hand, the usefulness of the calculi underlying \acp{FM} is
directly proportional only to their completeness and~(traditionally)
expires with a precision of less than $1$.  In other words, we usually
try to avoid calculi that allow the derivation of false theorems.  On
the other hand, semi-formal pattern checkers have shown to have a
wide, sometimes unacceptable, range of precision and recall of their
findings.  The usefulness of practical \ac{FM}-based tools might lie
somewhere in the middle between classical calculi and bug finding
tools with poor precision/recall values.

\begin{threat}
  The main threat is the struggle of academia and tool vendors to
  provide adequate tools, suited to adapt to novel scientific
  insights, to be integrated with other tools, and to be maintained in
  a flexible and independent manner.  
\end{threat}

\begin{remedy}
  To reduce this threat, an improvement of education and strong
  incentives can play an important role
  here~(\cf\Cref{sec:tool-integration,sec:host-transf-cult}).
\end{remedy}

\begin{table*}[t]
  \caption{Overview of our \ac{SWOT} analysis \cite{Piercy1989-MakingSWOTAnalysis} of ``\acp{iFM} in practical 
    \ac{RAS} assurance'' 
    \label{tab:swot:ifm-in-ras}}
  \footnotesize
  \begin{tabularx}{\textwidth}{XX}
    \toprule
    \textbf{Method Strengths: \acp{iFM} raise the potential of \dots}
    \begin{itemize}
    \item improvement of \ac{RAS} models, specification of \ac{RAS} requirements,
      automation of \ac{RAS} verification~(\Cref{sec:strweak:reputation})
    \item early detection of systematic errors in \ac{RAS}
      designs~(\Cref{sec:strweak:reputation})
    \item helpful abstractions to inform engineers about critical \ac{RAS}
      properties~(\Cref{sec:strweak:expressivity})
    \item integration and coordination of several \acp{FM} to consistently
      reason about interrelated \ac{RAS} properties~(\Cref{sec:strweak:integration})
    \end{itemize}
    \textbf{Community Strengths: \ac{iFM} research can rely on \dots}
    \begin{itemize}
    \item many transfer re-entry points from former \ac{FM} case studies
      in industrial and academic labs~(\Cref{sec:strweak:transfer})
    \item many assurance practitioners who perceive \ac{FM} usefulness as
      positive~(\Cref{sec:strweak:reputation})
    \item research designs for comparative studies of \ac{FM}
      effectiveness~(\Cref{sec:strweak:evidence})
    \end{itemize}
    &
      \textbf{Method Weaknesses: \acp{FM} have suffered from \dots}
      \begin{itemize}
      \item being difficult to learn and apply, many assurance practitioners
        perceive ease of use of \acp{FM} as
        negative~(\Cref{sec:strweak:reputation})
      \item low effectiveness of formal models 
        because of the reality gap~(\Cref{sec:strweak:expressivity})
      \item fragile effectiveness and productivity in \ac{RAS}
        engineering~(\Cref{sec:strweak:evidence})
      \end{itemize}
      \textbf{Community Weaknesses: \ac{iFM} progress has been hampered by
      \dots} 

      \begin{minipage}[t]{\linewidth}
        \begin{itemize}[nosep,after=\strut]
        \item no agreed framework for integration of
          \acp{FM}~(\Cref{sec:strweak:integration})
        \item lack of convincing evidence of \ac{FM} effectiveness in \ac{RAS}
          engineering~(\Cref{sec:strweak:evidence})
        \item research ineffectively communicated in \ac{iFM}
          teaching/training~(\Cref{sec:strweak:reputation})
        \end{itemize}
      \end{minipage}
    \\\midrule
    \textbf{Key Opportunities for \ac{iFM} transfer and progress:}
    \begin{itemize}
    \item The desire for early removal of erroneous \ac{RAS} behaviour and
      model-based assurance~(\Cref{sec:desire:earlyremoval})
    \item The desire to learn from \ac{RAS} accidents and their root causes~(\Cref{sec:desire:understerror})
    \item The desire of \ac{RAS} assurance to be a mature discipline~(\Cref{sec:desire:maturity})
    \item The desire for adequate and dependable \ac{RAS}
      norms~(\Cref{sec:desire:problemstandards})
    \end{itemize}
    \vspace{6em}
    The analysis provided in this table is an enhancement of the
    general analysis in \Cref{tab:swot:ifm-in-dse} in 
    \Cref{sec:swot-general}.
    &
    \textbf{Method Threats: \ac{iFM} research is threatened by \dots}
    \begin{itemize}
    \item misconceptions of semantic unification in \ac{RAS}
      assurance~(\Cref{sec:misp-unif})
    \item \acp{iFM} not scaling up to industry-size \acp{RAS}~(\Cref{sec:erroneousproofs})
    \item faulty, tedious, or vacuous
      proofs~(\Cref{sec:erroneousproofs,sec:soundn-usef-autom})
    \item poor integration with \ac{RAS} engineering tools,
      processes, and education~(\Cref{sec:tool-integration,sec:erroneousproofs,sec:soundn-usef-autom})
    \end{itemize}
    \textbf{Transfer Threats: \ac{iFM} transfer is threatened by 
      \dots}
      
      \begin{minipage}[t]{\linewidth}
        \begin{itemize}[nosep,after=\strut]
        \item a lack of roboticists' education in
          \acp{iFM}~(\Cref{sec:erroneousproofs,sec:tool-integration})
        \item a lack of \ac{iFM} researcher engagement in transfer to \ac{RAS}
          practice~(\Cref{sec:host-transf-cult})
        \item a lack of comprehensive access to quality data from \ac{RAS}
          practice~(\Cref{sec:scepticism})
        \end{itemize}
      \end{minipage}
    \\\bottomrule
  \end{tabularx}
\end{table*}

\section{A Vision of Integrated Formal Methods for Assurance}
\label{sec:position}

The following discussion applies to many domains of dependability
assurance.  However, the complexity of robots and autonomous systems
forms a key opportunity for the progress of \ac{iFM} research and for
its successful transfer.  Accordingly, \Cref{tab:swot:ifm-in-ras}
summarises the discussion in
\Cref{sec:threats,sec:opportunities,sec:strengthsweaknesses} with an
interpretation into \ac{RAS} assurance practice.  Based on the strengths
and opportunities described in the
\Cref{sec:opportunities,sec:strengthsweaknesses}, we formulate our
vision in terms of working hypotheses:

\begin{enumerate}[(H1)]
\item From \Cref{sec:desire-model-based}: Software tools for
  the construction of arguments and production of evidence using
  \emph{\acp{iFM} can meet the challenge} of assuring \ac{RAS}
  safe.  Computer-assisted assurance cases supported by heterogeneous
  formal models will increase confidence in their sufficiency,
  and also aid in maintenance and evolution through
  modularisation of arguments and evidence.
\item From \Cref{sec:desire-model-based,sec:desire:maturity}: \acp{iFM}, in
  particular modern verification tools, will enable \emph{automation
    of the evidence gathering process}, and highlight potential
  problems when an assurance case changes or when an incident occurs.

\item From \Cref{sec:desire:earlyremoval,sec:desire:maturity}:
  There is \emph{no stable path to assured autonomy without
    the use of \acp{iFM}}.  Acceptable safety will be much more likely
  achieved with \acp{iFM} than without them.
\item From \Cref{sec:strweak:integration}: The success of
  \acp{iFM} depends on the ability to \emph{integrate a variety of \acp{FM}} for
  different aspects of \ac{RAS}~(\egs \ac{HCI},
  safety-security interaction, missing human fallback,
  environment/world modelling, uncertain prediction/behaviour), which
  is not currently possible.
\item From \Cref{sec:strweak:expressivity,sec:strweak:integration}:
  Sophisticated techniques for \emph{model integration and
    synchronisation are necessary} to support \ac{MDE} with \acp{iFM}.  This
  way, \acp{iFM} will make it easier to express consistent \ac{RAS} models
  covering all relevant aspects, make their modelling assumptions
  explicit, and improve future assurance practices.

\item From
  \Cref{sec:strweak:reputation,sec:strweak:transfer,sec:strweak:evidence,sec:host-transf-cult}:
  \acp{iFM} can be beneficial in the short term.  However, an important
  engineering principle is to be
  conservative~\cite{Perlis1969-IdentifyingDevelopingCurricula} and,
  therefore, not to change procedures unless there is compelling
  evidence that \acp{iFM} are effective.  Such evidence can be delivered
  through \emph{empirical
    research}~\cite[\egs][]{Pfleeger1997-Investigatinginfluenceformal,Sobel2002,Woodcock2009,Jeffery2015} %
  and collaboration of academia and industry.  That evidence
  \emph{is required to re-evaluate research and foster research
    progress and transfer}.
\item %
  From \Cref{sec:desire:understerror}: The demonstration of cost
  effectiveness in addition to technical effectiveness of new
  \acp{iFM} is necessary to justify further research.
\item From \Cref{sec:desire:problemstandards}: Norms are a lever of
  public interest in dependability~\cite{Feitelson2019-TonysLaw}.
  Current norms seem to deviate from the state of the art and may fail
  to guarantee product certification procedures that satisfy the
  public interest.
\end{enumerate}

\begin{figure}
  \centering
  \footnotesize
  \begin{tikzpicture}
    [dim/.style={align=flush left},
    tit/.style={align=left,anchor=south west},
    itm/.style={align=center,minimum height=3em,minimum
      width=2cm,fill=gray!20}, lab/.style={align=center}]

    \node[itm] (ifm) at (0,0) {integrated\\Formal Methods (\acp{iFM})};
    \node[itm] (efm) at ($(ifm)+(-8em,-10em)$) {Empirical
      Research\\of Formal Methods};
    \node[itm] (ras) at ($(ifm)+(8em,-10em)$) {Assurance Practice in\\Robots
      and Autonomous\\Systems (\acs{RAS})};

    \draw[arrows={-latex},thick,tips=proper]
    (ifm) edge[above,sloped,bend left] node[lab] {(H1-3) have potential\\
      to improve} (ras)
    (efm) edge[above] node[lab] {observes} (ras)
    (ras) edge[below,sloped] node[lab] {(H4-5) pose chal-\\lenges to} (ifm)
    (efm) edge[above,sloped,bend left] node[lab] {(H6-8) examines the\\effectiveness of} (ifm)
    ;
  \end{tikzpicture}
  \caption{Progress of research on integrated formal methods through
    transfer into and improvement of assurance practice of robots and autonomous systems
    \label{fig:overview}
  }
\end{figure}
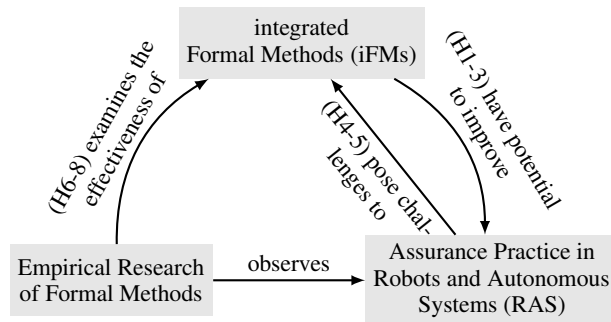

\Cref{fig:overview} assigns these hypotheses to the relationships
between foundational and transfer-directed \ac{iFM} research by
example of the \ac{RAS} domain.  Overall, we believe that \acp{iFM}
have great potential and can improve assurance but practitioners do
not use them accordingly.

\begin{opportunity}
  We could take and enhance credible measures to convince assurance
  practitioners of our results and effectively transfer these results.
  For this to happen, we have to answer further research questions.
\end{opportunity}

\section{Empirical, Applied, and Foundational Research}
\label{sec:agenda}

Based on the aforementioned working hypotheses, we state several
objectives for foundational and transfer-directed \ac{iFM} research,
formulate research questions, and show our expectations on desirable
outcomes of such research.

\subsection{Research Objectives and Tasks}
\label{sec:researchagenda}

To validate and transfer our research results, we need to
\begin{itemize}
\item evaluate how assurance case construction and management for
  certifiable \acp{RAS} can be improved by
  \acp{iFM}~\cite{Gleirscher2019-EvolutionFormalModel}.
\item debunk or justify arguments against the use of \acp{FM} or \ac{FM}-based
  tools in \ac{RAS} assurance.
\item foster a culture of successful \ac{FM} research transfer to
  industries performing \ac{RAS} assurance.
\end{itemize}
Taking an \ac{iFM} foundational point of view, we need
\begin{itemize}
\item foundational research on the integration and unification of \acp{FM} to
  tackle the complexity of \acp{RAS}~\cite{Luckcuck2018}.
\item a unified semantic foundation for the plethora of notations in
  \ac{RAS} assurance, to enable method and tool integration.  There
  are a number of promising research directions still being
  investigated~\cite{Hoare1998-UnifyingTheoriesProgramming,Rosu2010}.
\end{itemize}
Taking an evidence-based point of view, as already highlighted in 1993
by~\citet{Bowen1993-Safetycriticalsystems}, we need to
\begin{itemize}
\item understand the difference between the state of assurance
  practice and assurance research.
\item understand in which ways current \ac{RAS} assurance practices fail,
  and suggest effective approaches from assurance research.
  In this way, we can be sure that assurance practice is equipped with
  state-of-the-art assurance technology for defence against potential
  liability claims, and that assurance practitioners do not fail in
  fulfilling their obligations.
\item understand how results from assurance research can be validated
  to be sure that research follows promising directions with high
  potential of success in assurance practice.
\end{itemize}
Based on that, we need to 
\begin{itemize}
\item set concrete directions for empirical \ac{FM} research in \ac{RAS}
  assurance.
\item train \ac{FM} researchers in applying empirical research designs in
  their work on rigorous assurance cases.
  \citet{Woodcock2009} corroborated this objective by
  saying that ``formal methods champions need to be aware of the need
  to measure costs''.
\item avoid biases as found in various branches of scientific
  research.  \Eg in the social and biomedical sciences, researchers
  identified such biases through meta-analyses and suggested
  measures for bias
  avoidance~\cite[\egs][]{Fanelli2017-Metaassessmentbias,
    Franco2014-Publicationbiassocial}.
\item increase the level of evidence of \ac{FM} research to level 2
  according to the hierarchy
  in~\cite[Tab.~2]{Goues2018-BridgingGapResearch}.
\item avoid knowledge gaps about whether (a) \ac{RAS} practice is keeping
  up with the state of the assurance art, and (b) whether recent academic
  or industrial research is going in the right direction.
  In this way, we can be sure that we do our best to inform and
  serve the society.
\end{itemize}
Using appropriate research designs, we need to
\begin{itemize}
\item invite the \ac{RAS} industry to enhance their efforts in engaging
  with recent \ac{iFM} research.
\item foster goal-oriented interaction (a) between assurance
  practitioners and researchers and (b) between \ac{FM} researchers 
  and assurance researchers.
  In this way, we can be sure to do everything to keep researchers up
  to date with respect to practical demands.
\item join the \ac{FM} research and applied assurance research
  communities~(\Cref{fig:refinedapproach} on
  page~\pageref{fig:refinedapproach}), both vital for the progress and
  transfer of assurance research into \ac{RAS} assurance practice.
  This way, we can be sure to foster necessary knowledge transfer
  between these two communities.
\item summarise achievements in practical applications of \acp{iFM}
  for constructing assurance cases.
\item suggest improvements of curricula for \ac{RAS} assurance.
\item guide the \ac{RAS} industry in process improvement, training,
  and tool support.
\item guide vendors of \ac{FM}-based assurance tools to assess and improve
  their tools and services.
\end{itemize}

\subsection{Some Research Questions addressing these Objectives}
\label{sec:research-questions}

The research questions below are relevant for \acp{FM} in general.  We
consider these questions as crucial to be answered for \ac{RAS} assurance
to address the aforementioned objectives:
\begin{enumerate}[(Q1)]
\item What is the true extent of computer-related accidents up to
  2019~\cite{MacKenzie1994-Computerrelatedaccidental}?  What would
  these figures mean for the \ac{RAS} domain?
\item Does the use of formalism detect severe errors to a larger
  extent than without the
  use of formalism~\cite{Pfleeger1997-Investigatinginfluenceformal,Sobel2002}? 
\item Does the use of formalism detect severe errors earlier than
  without using formalism? 
\item Why would such errors be a compelling argument for the use of
  \acp{FM}? 
\item Apart from error avoidance and removal, which other benefits of
  \acp{iFM} in practice are evident and can be utilised for method
  trade-offs?
\item Which criteria play a central role in measuring \ac{iFM}
  effectiveness?  Particularly, how can \ac{FM}-based tools be used at
  scale?
\item How would Commercial-off-the-Shelf and
  System-Element-out-of-Context verification by \acp{iFM} pay off?
\item Which hurdles need to be overcome to use \acp{iFM} in practice
  to the maximum benefit?
\item How do we know when these hurdles are actually overcome?
\item How can \acp{FM}~(from different disciplines) be used
  together~(\acp{iFM}, unification)?
\item How can \acp{FM} be used to assure systems~(\egs computer vision
  in road vehicles) involving \ac{AI} techniques like
  machine learning and deep neural networks?
\item How can \acp{FM} be integrated into assurance cases to support
  certification against international safety and security standards?
\item How can we best combine formal and informal methods?  \Eg how
  can we deal with the issue of the validity of specifications that
  proofs rely on \cite{Jones2003-earlysearchtractable}.
\item How can we best present formal requirements, evidence, and artefacts in an
  assurance case?
\item How can empirical research help in successfully demonstrating
  the capabilities of \acp{iFM} for rigorous and certifiable autonomy
  assurance?
\end{enumerate}
This list of research questions can easily be extended by further more
detailed empirical questions from the settings discussed in 
\cite[Sec.~4.4]{Jones2011-EconomicsSoftwareQuality}.

\subsection{Some Envisaged Research Outcomes}
\label{sec:topics}

Our vision of \emph{rigorous \ac{RAS} assurance} implies foundational \ac{iFM}
research to result in
\begin{itemize}
\item novel semantic frameworks unifying best practice methods,
  models, and formalisms established in \ac{RAS}
\item new concepts for \ac{iFM}-based development environments
\item new computational theories to support formal modelling and
  verification of \ac{RAS}
\item evaluations of
  \begin{itemize}
  \item assurance tools, languages, frameworks, or platforms used in
    practice regarding their support of \acp{iFM}
  \item the integration of \acp{iFM} into modelling and programming
    techniques, assurance methods, and assurance processes
  \item languages for linking informal requirements with evidence from
    \acp{iFM}
  \item (automated) abstraction techniques used in assurance and
    certification
  \end{itemize}
\item opinions, positions, and visions on \ac{FM} integration and
  unification for rigorous practical assurance.
\end{itemize}
Our vision of \emph{rigorous \ac{RAS} assurance} implies applied and
empirical \ac{iFM} research to result in
\begin{itemize}
\item comparisons of
  \begin{itemize}
  \item projects applying \acp{iFM} in assurance practice with similar practical
    projects applying non-\ac{iFM} approaches
  \item \ac{iFM}-based~(embedded software) assurance with assurance
    approaches in traditional engineering disciplines
  \end{itemize}

\item checklists, metrics, and benchmarks~(for and beyond tool performance)
  for
  \begin{itemize}
  \item the evaluation and comparison of \ac{iFM}-based assurance
    approaches~(\egs confidence level)
  \item relating error removal and incident root cause data~(\egs
    efficiency and effectiveness in removal of severe errors or in
    avoidance of severe accidents,
    \cf\cite{MacKenzie1994-Computerrelatedaccidental})
  \item usability and maturity assessment of \acp{iFM}~(\egs abstraction
    effort, proof complexity, assurance case complexity, productivity)
  \item the evaluation of \ac{FM} budget
    cases~(\cf\cite{Darbari2018-budgetcaseformal} in electronic
    hardware development).
  \end{itemize}

\item experiences in or surveys~(\egs systematic mappings and reviews
  of assurance case research, interview studies with assurance
  practitioners, \cf\Cref{sec:excurs-relat-betw}) of
  \begin{itemize}
  \item \ac{iFM} transfers and applications~(\egs case studies in assurance
    and certification projects)
  \item challenges, limitations/barriers, and benefits of
    \acp{iFM} in assurance and certification projects,
  \end{itemize}
  
\item research designs~(\egs for controlled field experiments) for the
  practical validation of \acp{iFM} in assurance and certification projects

\item opinions, positions, and visions on future research, education,
  and training in the use of \acp{iFM} in assurance and certification.
\end{itemize}

\section{Summary}
\label{sec:conlusions}

Along the lines of \citet{Hoare2009-verifiedsoftwareinitiative}, we
analysed strengths, weaknesses, opportunities, and threats to
determine the potential of integrated formal methods to improve the
practice of dependability assurance.  Emphasising robots and
autonomous systems as an area in the spotlight of dependability
assurance, we express our expectations of research progress and
transfer.  From these expectations, we derived a research and research
transfer agenda with the objective of
\begin{inparaenum}[(i)]
\item enhancing the foundations of integrated formal methods,
\item collecting evidence on the effectiveness of integrated formal
  methods in practice,
\item successfully transferring integrated formal methods into the
  assurance practice, with a short-term focus on robots and autonomous
  systems, and
\item fostering research progress, education, and training from the
  results of this transfer effort.
\end{inparaenum}

\paragraph{Acknowledgements}

This work is partly supported by the Deutsche
Forschungsgemeinschaft~(DFG) under the Grant no.~381212925 and the
EPSRC projects CyPhyAssure,\footnote{CyPhyAssure Project:
  \url{https://www.cs.york.ac.uk/circus/CyPhyAssure/}.} grant
reference EP/S001190/1, and RoboCalc, grant reference EP/M025756/1.
We are deeply grateful to Michael Butler, Ana Cavalcanti, John
Fitzgerald, Cliff Jones, and Peter Gorm Larsen for helpful feedback
on previous versions of this manuscript and for their genuine
support in shaping the direction of this work.

\footnotesize
\bibliography{references}
\normalsize

\newpage
\appendix
\section*{\appendixname}
\section{Studying Formal Method Practitioners: A Ballpark Figure}
\label{sec:excurs-relat-betw}

In this section, we make a brief excursion to the relationship between
\acp{FM} and \ac{FI}, closing with a rough estimate of the size of the
population of \ac{FM} users and the meaning of such an estimate for
the empirical study of that population.  Such studies include
cross-sectional and longitudinal surveys based on questionnaires or
semi-structured interviews.

\paragraph{Formal Inspection versus Formal Methods}

\Ac{FI} encompasses a variety of techniques~(\egs peer reviews,
walk-throughs) where critical process artefacts~(\egs program code)
are checked (\egs manually or using software tools) against a variety
of criteria~(\egs checklists), usually by a group of independent
qualified engineers.  For the sake of simplicity of the following
discussion, we assume that \acp{FM} can be seen as a particularly
rigorous variant of \ac{FI} where formal specifications serve as a
particular way of formulating checklists.

We compare the use of \acp{FM} with the use of \ac{FI}.  According to
\citet[Sec.~4.4]{Jones2011-EconomicsSoftwareQuality}, formal
inspections are used in more than 35\% of commercial defence, systems,
and embedded software projects, and \acp{FM} are estimated to be applied in
less than 1\% of overall commercial software engineering projects.  To
get an idea of this coverage data, we perform an analysis of the
global embedded software market based on other global software market
indicators in \Cref{tab:swmarket}.  We found estimates of systems and
software professionals world-wide and estimates of annual US business
values.%
\footnote{See \url{https://en.wikipedia.org/wiki/Software_industry}
  and
  \url{https://softwareengineering.stackexchange.com/questions/18959/what-proportion-of-programming-is-done-for-embedded-systems}.}

A uniform distribution would entail roughly 37000 USD/year per person
in the general software domain and 10000 USD/year per person in the embedded
software domain.  Clearly, geographically strongly differing salaries and
part-time engagement rule out a uniform distribution, yet providing
figures helpful for our purposes.

Next, we apply the following proportions: From a worldwide population
of around 18.5 million software developers in
2014~\cite{EvansData2018}, about 19\% live in the US, 10\% in China,
9.8\% in India, 36\% Asia/Pacific region, 39 \% live in Europe, the
Middle East, and Africa; and 30\% in the Americas.%
\footnote{See
  \url{https://adtmag.com/Blogs/WatersWorks/2014/01/Worldwide-Developer-Count.aspx}.}
The \emph{design to quality assurance}~(\ies verification and test)
\emph{cost ratio} is observed to be approximately $30:70$.%
\footnote{See \url{https://www.slideshare.net/pboulet/socdesign}.}
About 20\% of embedded software personnel are quality assurance
engineers~(\ies test, verification, or validation engineers).%
\footnote{See
  \url{https://de.slideshare.net/vdcresearch/searching-for-the-total-size-of-the-embedded-software-engineering-market}.}

\begin{table*}
  \caption{
    Data for the estimation of the size of the formal inspection and
    formal method market
    \label{tab:swmarket}
  }
  \footnotesize
  \begin{tabularx}{1.0\linewidth}{p{3.5cm}|XXXXllX} 
    \toprule
    \emph{Global market/project indicators} [Unit]
    & \emph{Professional engineers / developers} [million]
    & \emph{Ann.~business value} [billion USD/year]
    & \emph{Quality assurance personnel} [million]
    & \emph{\acs{QA} business value} [billion USD/year]
    & \emph{\acs{FI}} [\%]
    & \emph{\acs{FM}} [\%]
    & \emph{Devices} [billion / year]
    \\\midrule
    General IT hardware and devices (incl.~personal computers)
    & 
    &
      2018: 689
    & 
    & 
    &
    & 
    & 2010: 10
    \\\midrule
    Embedded systems (hardware, software, connected embedded devices) in all domains 
    &
      2014: 1.2 %
    & 2009: 88\newline
      2018: (161)$^a$
    &
    & 
    & (35)
    &
    & 2010: 9.8
    \\\midrule
    Industrial embedded systems
    & 
    &
    &
    & 
    & 
    &
    & 2016: 2
    \\\midrule
    Defence, systems, and embedded commercial software engineering
    & 
    & 
    & 
    & 
    & 2011: 35
    &
    \\\midrule
    Embedded software 
    & 2014: 0.36
    & 2009: 3.4\newline
      2018: (10)
    & \textbf{2010: (0.072)}
    & \textbf{2010: (2.38)}
    & 
    &
    \\\midrule
    General software (overall commercial software engineering, development)
    & 2014: 18.5\newline      
      2018: 23\newline
      2019: 23.9 %
    &
      2013: 407\newline
      2018: (515)
    &
    &
    & (4.6)
    & 2011: 1
    & 
    \\\bottomrule
    \multicolumn{8}{p{13cm}}{$^a$The numbers in parentheses include
    estimates for 2018 based on the other numbers and corresponding
    average growth rates.} 
  \end{tabularx}
\end{table*}

The estimates in \Cref{tab:swmarket} suggest that around 2\% of the
overall pure software market are allocated to the embedded pure software
market.  $35\%$ coverage of formal inspection in about 13.5\% of the
overall software market~($161/(689+515) = 0.134$) would result in
roughly 4.7\% coverage of all software projects by formal inspection
versus at most 1\% coverage by \acp{FM}.  However, from this data we can
hardly know whether rates of \ac{FM} use get close to or beyond 10\% in
high-criticality systems projects.

Assuming that in about 35\% of embedded software projects the quality
assurance personnel would use formal inspection and that in every
fifth~($1:4.7$) of such projects formal methods would be used, the
current population of regular practical \ac{FM} users would globally amount
to about 5040~($= 72,000 * 0.34 * 0.20$) persons.  Note that these
numbers are rough estimates.  However, we believe their order of
magnitude is realistic.  Given that these persons would on
average earn about 100,000 USD/year each, we would speak of an annual
business value of around USD~504 million.

\paragraph{Empirical Studies of \ac{FI} and \ac{FM} Practitioners}

Importantly, from these data we can determine minimum sample sizes for
surveys.  \Eg assume we want to have 95\% confidence in our test
results and are fine with a confidence interval of $\pm$7\%.  Then,
for \emph{regular practical \ac{FM} users}, a population of the size
of 5040 persons would require us to sample 189 independent data
points~(\egs questionnaire responses).  The population of
\emph{regular practical \ac{FI} users}, 25200~($=72,000*.35$) would
imply a minimum sample size of 194.

For any sample of such size, any survey has to argue why the sample
\emph{represents} the population.  This step depends on the
possibilities given during the sampling stage.  Obviously, reaching
out to 189 out of 5040 persons whose locations might be largely
unknown is an extremely difficult task that might only be tackled in
terms of a global group effort among researchers and practitioners.
Given that this challenge can be tackled, these figures determine the
type of surveys required for the collection of confident evidence.
Moreover, these figures limit the manual effort to perform and repeat
valuable qualitative studies to a reasonable amount.

\section{Formal Methods in Dependable Systems Engineering}
\label{sec:swot-general}

\Cref{fig:refinedapproach} on page~\pageref{fig:refinedapproach}
depicts our research setting for this survey.  There, we describe
\acl{iFM} as an enhancement of \aclp{FM} and the assurance of
\aclp{RAS} as a special field of dependable systems engineering.
Because many of the discussed issues are of a more general nature, we
first prepared a \ac{SWOT} analysis of \acp{FM} in dependable systems
engineering in \Cref{tab:swot:ifm-in-dse}.  Then, we refined this
analysis in \Cref{tab:swot:ifm-in-ras} to accommodate specifics of
\aclp{iFM} for \acl{RAS}.

\begin{table*}
  \caption{Overview of our general \ac{SWOT} analysis (according
    to~\citet{Piercy1989-MakingSWOTAnalysis}) of ``formal methods in
    dependable systems engineering''
    \label{tab:swot:ifm-in-dse}}
  \footnotesize
  \begin{tabularx}{\textwidth}{XX}
    \toprule
    \textbf{Method Strengths:}
    
    \begin{minipage}[t]{\linewidth}
      \begin{itemize}
      \item Improvement of modelling precision, requirements clarity,
        verification confidence
      \item High error detection effectiveness, early error removal
      \end{itemize}
      \textbf{Community Strengths:}
      \begin{itemize}
      \item Many transfer re-entry points from former case studies with
        industry 
      \item Many dependable systems practitioners perceive \ac{FM} usefulness as positive
      \end{itemize}
    \end{minipage}
    &
    \textbf{Method Weaknesses:}
    \begin{itemize}
    \item Difficult to learn and apply, many dependable systems practitioners
      perceive ease of use of \acp{FM} as negative
    \item Fragile effectiveness and productivity 
    \end{itemize}
    \textbf{Community Weaknesses:}
    \begin{itemize}
    \item Lack of compelling evidence of \ac{FM} effectiveness
    \item Ineffectively communicated in teaching and training
    \end{itemize}
    \\\midrule
    \textbf{Key Opportunities for \ac{FM} Transfer and Research:}
    \begin{itemize}
    \item Desire for early removal of severe errors (\Cref{sec:desire:earlyremoval})
    \item Desire to learn from accidents and their root causes (\Cref{sec:desire:understerror})
    \item Desire to be a mature discipline~(\Cref{sec:desire:maturity})
    \item Desire for dependable norms~(\Cref{sec:desire:problemstandards})
    \end{itemize}
    &
      \textbf{Method Threats:}
      
      \begin{minipage}[t]{\linewidth}
        \begin{itemize}
        \item Lack of method scalability
        \item Faulty, tedious, or vacuous proofs
        \item Lack of user education
        \item Poor tool integration, legacy tools and processes
        \end{itemize}
        \textbf{Transfer Threats:}
        \begin{itemize}
        \item Lack of researcher engagement in \ac{FM} transfer
        \item Lack of access to comprehensive high-quality data
        \end{itemize}
      \end{minipage}
    \\\bottomrule
  \end{tabularx}
\end{table*}

 \end{document}